%% file: main.tex
\pdfoutput=1
\documentclass[fleqn,usenatbib,usedcolumn]{mnras}

\usepackage{graphicx,amssymb,hyperref}
\usepackage{color}

\usepackage[fleqn]{amsmath}
\usepackage{abbrev}
\usepackage{siunitx}
\usepackage{gensymb}
\usepackage{booktabs,tabularx}
\usepackage{flushend}
\usepackage{xspace}
\usepackage{xcolor}
\usepackage{makecell}
\usepackage{orcidlink}

\usepackage[shortlabels]{enumitem}
\usepackage{subcaption}
\setlist[itemize]{leftmargin=5pt}

\def\mnras{MNRAS}
\def\apj{ApJ}

\def\simlt{\lower.5ex\hbox{$\; \buildrel < \over \sim \;$}}
\def\simgt{\lower.5ex\hbox{$\; \buildrel > \over \sim \;$}}
\def\etal{{\it et al.}}

\def\mpc{\mathrm{\, Mpc}}
\def\msun{\mathrm{\, M_\odot}}
\def\kms{\mathrm{\, km \, s^{-1}}}

\def\transpose{^\intercal}
\def\xicgd{\xi_\mathrm{cg}^{s,\mathrm{d}}}

\defcitealias{2012ApJ...761....1W}{W12}
\defcitealias{2019MNRAS.488.3646R}{R19}
\defcitealias{2021MNRAS.500.2627H}{H21}

\defcitealias{2013MNRAS.431.3319Z}{ZW13}
\newcommand{\zw}{\citetalias{2013MNRAS.431.3319Z}\xspace}

\newcommand{\subfind}{\textsc{subfind}\xspace}
\newcommand{\george}{\textsc{george}\xspace}

\def\gs{\mathrel{\raise1.16pt\hbox{$>$}\kern-7.0pt \lower3.06pt\hbox{{$\scriptstyle \sim$}}}}         
\def\ls{\mathrel{\raise1.16pt\hbox{$<$}\kern-7.0pt \lower3.06pt\hbox{{$\scriptstyle \sim$}}}}   

\newcommand{\vect}[1]{\boldsymbol{#1}}

\def\mrm{\mathrm}
\def\hgpc{h^{-1}\mathrm{Gpc}}
\def\hmpc{h^{-1}\mathrm{Mpc}}
\def\hkpc{h^{-1}\mathrm{kpc}}
\def\hmsol{h^{-1}M_\odot}
\def\hmsun{h^{-1}M_\odot}
\def\vr{v_{r}}
\def\vrc{v_{r, c}}
\def\fvir{f_{\mathrm{vir}}}
\def\sv{\sigma_{\mathrm{vir}}}
\def\sr{\sigma_{\mathrm{rad}}}
\def\st{\sigma_{\mathrm{tan}}}
\def\s0{\sigma_{0}}
\def\dof{\mathtt{dof}}
\def\rp{r_p}
\def\rpi{r_\pi}
\def\fr0{\bar{f}_{\mrm{R}0}}

\newcommand{\be}{\begin{equation}}
\newcommand{\ee}{\end{equation}}
\newcommand{\ba}{\begin{eqnarray}}
\newcommand{\ea}{\end{eqnarray}}

\newcommand{\Om}{\Omega_\mathrm{m}}

\newcommand{\xicg}{\xi_\mathrm{cg}^s}

\DeclareGraphicsExtensions{.pdf,.png}

\title[The redshift-space cluster-galaxy correlation function]{Modelling the redshift-space cluster--galaxy correlation function on Mpc scales with emulation of the pairwise velocity distribution}

\author[A.\ Robertson \etal]{Andrew Robertson$^1$\thanks{e-mail: {\tt andrew.a.robertson@jpl.nasa.gov}}\orcidlink{0000-0002-0086-0524}, Eric Huff$^1$\orcidlink{0000-0002-9378-3424}, Katarina Markovi\v{c}$^1$\orcidlink{0000-0001-6764-073X} and Baojiu Li$^2$\orcidlink{0000-0002-1098-9188}
\\$^1$Jet Propulsion Laboratory, California Institute of Technology, 4800 Oak Grove Drive, Pasadena, CA 91109, USA
\\$^2$Institute for Computational Cosmology, Department of Physics, Durham University, South Road, Durham, DH1 3LE, UK\\
}

\begin{document}

\maketitle

\label{firstpage}

\begin{abstract}

We present a method for modelling the cluster-galaxy correlation function in redshift-space, down to $\sim$ Mpc scales. The method builds upon the so-called Galaxy Infall Kinematics (GIK) model, a parametric model for the pairwise velocities of galaxies with respect to nearby galaxy clusters. We fit the parameters of the GIK model to a suite of simulations run with different cosmologies, and use Gaussian Processes to emulate how the GIK parameters depend upon cosmology. This emulator can then be combined with knowledge of the real-space clustering of clusters and galaxies, to predict the cluster--galaxy correlation function in redshift space, $\xicg$. Fitting this model to an observed $\xicg$ enables the extraction of cosmological parameter constraints, and we present forecasts for a DESI-like survey. We also perform tests of the robustness of our constraints from fitting to mock data extracted from $N$-body simulations, finding that fitting to scales $\lesssim 3 \, \hmpc$ leads to a biased inference on cosmology, due to model misspecification on these scales. Finally, we discuss what steps will need to be taken in order to apply our method to real data.

\end{abstract}

\begin{keywords}
large-scale structure of Universe - galaxies: haloes - galaxies: clusters: general - cosmology: theory
\end{keywords}

\section{Introduction}

Understanding the cause of the accelerated expansion of the late-time Universe is one of the primary goals of 21st century cosmology \citep{2008ARA&A..46..385F}. While current observational data appears to be consistent with the expansion being driven by a cosmological constant, $\Lambda$, an exciting alternative is that the accelerated expansion is pointing towards gravity behaving in a different manner from the predictions of General Relativity (GR) \citep{2016ARNPS..66...95J}. There are numerous modified gravity theories, which -- as well as providing potential explanations for the accelerated expansion -- generally predict regimes in which there are enhancements to the total gravitational force, above that expected with GR.

Because gravity on the scale of the solar system is known to be consistent with GR to high precision, these additional forces must somehow be \emph{screened} within the solar system. Screening mechanisms typically lead to the enhancements to gravity disappearing in high-density, or deep gravitational potential environments \citep{2013CQGra..30u4005B}. In this context, the infall of galaxies onto galaxy clusters provides an interesting test of these modified gravity theories, because galaxies will transition from being unscreened to screened as they fall into the cluster potential \citep{2014MNRAS.445.1885Z}.

Clusters and their masses are also an interesting area for current exploration, because of current tensions between measurements of cosmological parameters from cluster abundance studies and measurements made using other methods. For example, the Dark Energy Survey's Year 1 Results indicated that the number and/or masses of galaxy clusters disagreed with that expected from the cosmology fit to combined cosmic shear, galaxy--galaxy lensing and photometric galaxy clustering \citep{2020PhRvD.102b3509A}. To quantitatively explain these results, the masses inferred through weak lensing for DES's lowest-richness clusters would have to be 30--40\% lower than expected from the DES 3x2pt cosmology, or else the richness-selected sample of clusters must be highly ($\sim$50\%) incomplete. 

The cluster--galaxy cross-correlation function in redshift space, $\xicg$, is an observable quantity that depends both on the clustering of galaxies around galaxy clusters, and on the relative velocities of galaxies with respect to nearby clusters. These velocities in turn depend upon the mass distribution of clusters, as well as the theory of gravity. $\xicg(r_p, r_\pi)$ measures the excess probability (above that expected for randomly distributed clusters and galaxies) of measuring cluster galaxy pairs with an inferred separation along the line-of-sight of $r_\pi$, and a separation perpendicular to this of $r_p$. Cluster--galaxy pairwise velocities get imprinted into $\xicg(r_p, r_\pi)$ because the line-of-sight separations between cluster--galaxy pairs are inferred from differences in their redshifts, which -- as well as depending upon the true line-of-sight separations -- receive a contribution from the line-of-sight components of their peculiar velocities. The effects of peculiar velocities on the observed clustering signal are known as redshift space distortions (RSD).

If we wish to use RSD in the cluster--galaxy correlation function to probe cluster mass distributions as well as modifications to GR, then we need to include data from the small scales on which the cluster potential makes a significant contribution to the galaxy velocities. Clustering on these small scales also contains a  large fraction of the statistical constraining power from current and future surveys, making the accurate modelling of redshift-space clustering on these scales a crucial ingredient for extracting the maximum amount of information. For example, for a BOSS-like survey, the constraints on the growth rate of structure from $r < 30 \, \hmpc$ scales alone are nearly a factor of two tighter than those from the fiducial BOSS analysis of redshift-space clustering using perturbation theory on larger scales \citep{2019ApJ...874...95Z}.


Given the ever increasing volume and resolution of $N$-body simulations, one approach to modelling galaxy clustering on small scales is to directly use the results from a simulation with a given cosmology as the model prediction for that cosmology. In practice, simulations are still too time consuming to run one for every likelihood evaluation in a cosmological likelihood analysis. However, assuming that the simulated data vector varies smoothly as the cosmological parameters are varied, some sort of interpolation scheme can be used to predict the data vector at an arbitrary cosmology, from the data vector calculated from simulations run at other cosmologies.

This procedure is known as \emph{emulation} \citep[see][for a few different examples in cosmology]{2019MNRAS.484.5509E, 2019ApJ...874...95Z, 2022MNRAS.515..871Y}. Emulating some cosmological observable from simulations requires that we trust the simulations to reliably predict the observable. It also requires simulation volumes comparable with or larger than observed survey volumes, so that noise in the simulations does not dominate the error budget -- although there have been recent developments to suppress the variance in simulation predictions \citep[e.g.][]{2016MNRAS.462L...1A, 2022JCAP...10..036M, 2022JCAP...09..059K}.

Here, we explore an alternative to directly emulating observables, which is to use a physically motivated model, the parameters of which we expect to depend systematically on the values of the cosmological parameters. We can then build an emulator for the parameters of the model, rather than for the observed data vector itself. This approach has some advantages and disadvantages over straight emulation of the data vector (discussed in Section~\ref{sect:pros_cons_vs_direct_xicg}). In short, a physical model can suppress noise and also aids with physical understanding, but opens the door to the potential for model-misspecification. We build such a model for the redshift-space cluster-galaxy correlation function, employing the galaxy infall kinematics (GIK) model from \citet[][\zw hereafter]{2013MNRAS.431.3319Z} to describe the distribution of cluster--galaxy pairwise velocities. We build an emulator for the parameters of the GIK model using the FORGE suite of $f(R)$ modified gravity simulations \citep{2022MNRAS.515.4161A}, and then demonstrate fitting this to mock observations of the cluster--galaxy correlation function in redshift-space.

This paper is structured as follows. In Section~\ref{sect:xi_cg_s} we provide some basic definitions useful in the rest of the paper. Then in Section~\ref{sect:gik} we introduce the GIK model from \zw, highlighting some small changes that we have made. In Section~\ref{sect:calc_xicg_with_GIK} we demonstrate how to combine the real-space correlation function with the GIK model in order to calculate the redshift-space correlation function. Then in Section~\ref{sect:GIK_param_emulator} we describe using simulations run with different cosmological parameters to build an emulator for the GIK model parameters. In Section~\ref{sect:fitting_to_mock_data} we demonstrate the efficacy of our model, by fitting it to mock data. In Section~\ref{sect:discussion} we outline some of the advancements required before our method could be applied to real observational clustering data, before concluding in Section~\ref{sect:conclusions}.

\section{The cluster--galaxy correlation function in redshift space}
\label{sect:xi_cg_s}

The cluster--galaxy correlation function in real-space, $\xi_\mathrm{cg}(r)$, is defined such that for a sample of galaxies with number density $n_\mathrm{g}$ and galaxy clusters with number density $n_\mathrm{c}$,
\begin{equation}
\mathrm{d}P_\mathrm{cg} = n_\mathrm{c} \, n_\mathrm{g} \left[ 1 + \xi_\mathrm{cg}(r) \right] \mathrm{d}V_\mathrm{c} \, \mathrm{d}V_\mathrm{g}
\end{equation}
is the joint probability of finding a galaxy cluster in the volume $\mathrm{d}V_\mathrm{c}$ and a galaxy in the volume $\mathrm{d}V_\mathrm{g}$, where the separation between the two (small) volumes is $r$. This probability (and hence $\xi_\mathrm{cg}(r)$) depends only on the distance between the two volumes (and not on the direction of their separation or their absolute positions) because of the assumptions of isotropy and homogeneity.

Calculating $\xi_\mathrm{cg}(r)$ is rather simple for the case of clusters and galaxies found within a cosmological simulation in a cubic (periodic) box. For some definition of `galaxy' and `galaxy cluster', we can count the number of galaxy--cluster pairs that have a separation between $r$ and $r + \mathrm{d}r$, which we call $\mathrm{d} N^\mathrm{sim}_\mathrm{cg}$. Assuming that the total simulation volume is $V$, the expected number of clusters is $n_\mathrm{c} V$. If we just consider one of these clusters, then the expected number of galaxies (if they were randomly distributed) a distance between $r$ and $r + \mathrm{d}r$ away from the cluster is $4 \pi r^2 \mathrm{d}r \, n_\mathrm{g}$, such that when accounting for all clusters we expect $\mathrm{d} N^\mathrm{rand}_\mathrm{cg} = 4 \pi r^2 \mathrm{d}r \, n_\mathrm{g} n_\mathrm{c} V$ pairs. Our estimate for the cluster--galaxy correlation function is then $\xi_\mathrm{cg}(r) = \mathrm{d} N^\mathrm{sim}_\mathrm{cg} / \mathrm{d} N^\mathrm{rand}_\mathrm{cg} - 1$. 

This same notion can be extended to the case of anisotropic clustering in redshift space, where the isotropic nature of galaxy clustering is broken by RSD. In an analogous manner to the real-space clustering case discussed above, we can count the number of cluster--galaxy pairs that have a separation with line-of-sight component in the range $\left[ \rpi ,  \rpi + \mathrm{d} \rpi \right]$, and a plane-of-sky component in the range $\left[ \rp ,  \rp + \mathrm{d} \rp \right]$, and compare this with the expectation in the case of randomly distributed clusters and galaxies to find $\xi^s_\mathrm{cg}(\rp, \rpi)$, where the superscript $s$ denotes that this correlation function is measured in redshift space. 

\subsection{The Indra simulations}
\label{sect:Indra}

We use the publicly available Indra suite of simulations \citep{2021MNRAS.506.2659F}. These will be useful for first elucidating some of the workings of the GIK model, and then later for calculating a covariance matrix for $\xi^s_\mathrm{cg}(\rp, \rpi)$. The Indra suite consists of 384 simulations, all run with the same cosmological parameters as one another, but with different random phases used for the initial conditions. The simulations are dark matter-only, and are of boxes with a side-length of $1 \, \hgpc$, with a particle mass of $7 \times 10^{10} \, \hmsol$. In total, each Indra simulation has 64 snapshots saved from the time of the initial conditions (at a redshift of $z=127$) down to $z=0$, but for simplicity we focus only on $z=0$ in this paper.

The Indra snapshots were analysed with the standard friends-of-friends (FOF) algorithm \citep{1985ApJ...292..371D}, after which the \subfind algorithm \citep{2001MNRAS.328..726S} was run, which calculates spherical overdensity quantities such as $M_{200}$ for each FOF group,\footnote{Centred on the most gravitationally bound particle in the FOF group, we define $r_{200}$ as the radius within which the mean enclosed density is $200 \times \rho_\mathrm{crit}$, with the mass enclosed within $r_{200}$ being $M_{200}$.} and also identifies gravitationally bound substructures within the FOF groups.

While there are numerous possible ways that one could populate a dark matter-only simulation with galaxies, and then also identify galaxy clusters, for now we will take a simplistic approach. We consider galaxy clusters to be all FOF groups with $M_{200}$ in some range, while galaxies are all \subfind subhalos with $v_\mathrm{max}$ above some threshold value. Here, $v_\mathrm{max}$ is the maximum value of the circular velocity, $v_\mathrm{circ}(r) = \sqrt{G M(<r)/ r}$, where $M(<r)$ is the mass enclosed within a radius $r$. Note that \subfind subhalos include the \emph{central} subhalo in each FOF group, so being tied to `subhalos' does not imply that our galaxies are necessarily substructures of something larger. Our galaxy assignment scheme can be thought of as subhalo abundance matching (SHAM), using $v_\mrm{max}$ as the \emph{abundance matching parameter} and with zero scatter \citep{2013ApJ...771...30R}.

For a cluster and galaxy at locations $\vect{r_\mathrm{c}}$ and $\vect{r_\mathrm{g}}$, and with peculiar velocities $\vect{v_\mathrm{c}}$ and $\vect{v_\mathrm{g}}$, respectively, we define $\vect{r} = \vect{r_\mathrm{g}} - \vect{r_\mathrm{c}}$ and $\vect{v} = \vect{v_\mathrm{g}} - \vect{v_\mathrm{c}}$ as the relative position and velocity of the galaxy with respect to the cluster. We also define the scalar cluster--galaxy separation,  $r = |\vect{r}|$. We denote a unit vector along the line-of-sight as $\vect{\hat{y}}$, with the line-of-sight separation between the cluster and galaxy being  $y = \vect{r} \cdot \vect{\hat{y}}$, and the separation perpendicular to this being $\rp = |\vect{r} - y \vect{\hat{y}}|$. The line-of-sight component of the relative velocity between the cluster and galaxy is $v_y = \vect{v} \cdot \vect{\hat{y}}$. The relative velocity along the line-of-sight changes the apparent line-of-sight separation between the galaxy and cluster to be $\rpi = y \, + \, (1+z) \, v_y / H(z)$, where the factor of $1+z$ accounts for the fact that we define $\rpi$ and $y$ in co-moving coordinates. 

\begin{figure}
        \centering
        \includegraphics[width=\columnwidth]{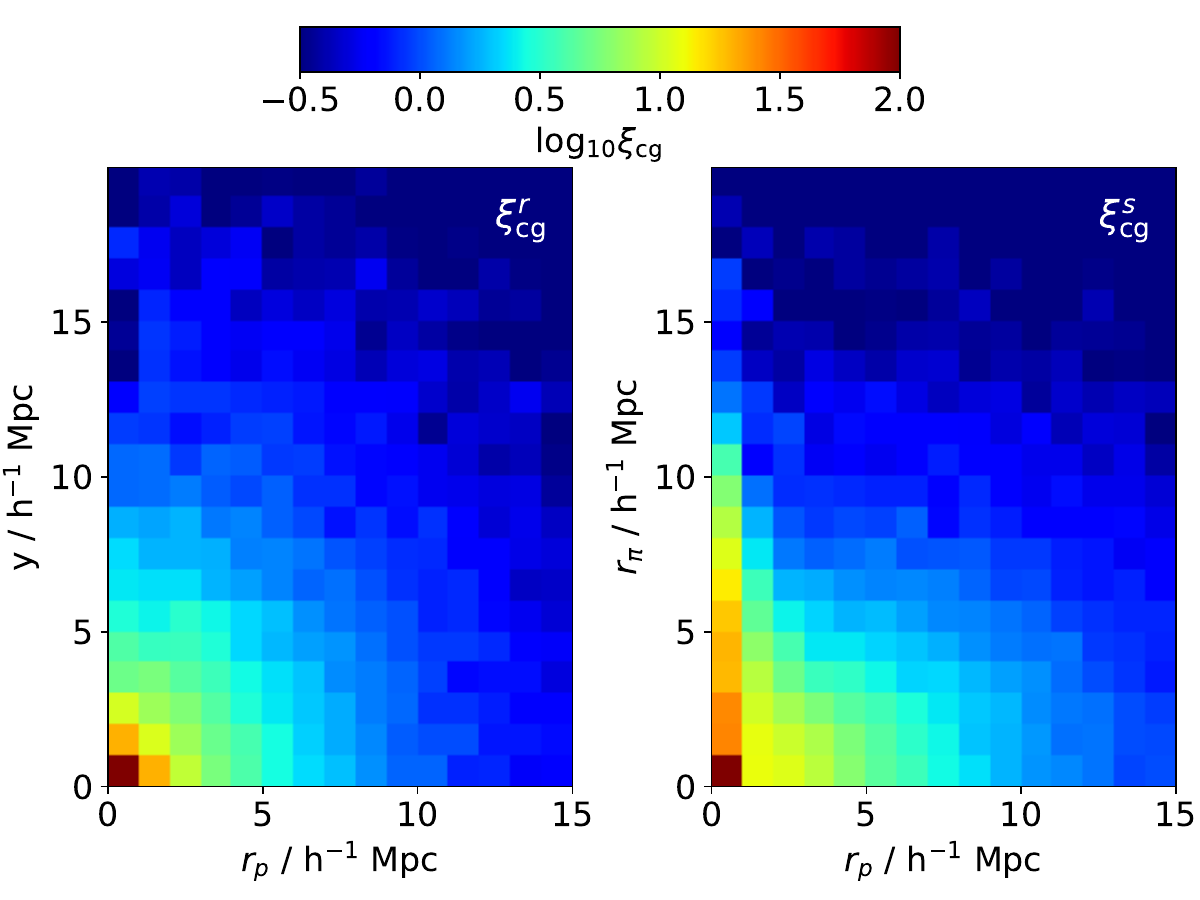}     
	\caption{The measured small-scale cluster--galaxy correlation function from the $z=0$ snapshot of an Indra simulation. The left panel shows the correlation function in real space, with $y$ the true line-of-sight separation between cluster--galaxy pairs, while the right panel uses the inferred line-of-sight separation (i.e. the separation in \emph{redshift space}), $\rpi$.}
	\label{fig:example_xi_cg}
\end{figure}

For a minimum galaxy $v_\mathrm{max}$ of $250 \, \kms$, and clusters with $14.1 < \log_{10} M_{200} / \hmsun < 14.2$, we plot the cluster--galaxy correlation function in Fig.~\ref{fig:example_xi_cg}. For this purpose we adopt one of the Cartesian axes of the simulation box as the line-of-sight, and then show the results both in real space (where the clustering is isotropic)\footnote{While the mean clustering signal is isotropic, the noise is not because (for example) a pixel at $(\rp = 10.5 \, \hmpc, y = 0.5 \, \hmpc )$ covers a larger volume than the pixel at $(\rp = 0.5 \, \hmpc, y = 10.5 \, \hmpc )$, such that the first of these has smaller fractional error due to Poisson counts.}
and in redshift-space, where \emph{Kaiser squashing} \citep{1987MNRAS.227....1K} enhances the clustering at intermediate-$\rp$ and low-$\rpi$, and the \emph{fingers of god} spread low-$r$ cluster--galaxy pairs along the line-of-sight.

This cluster definition leads to $\sim 3000$  clusters in each $(1 \, \hgpc)^3$ Indra volume, while the number density of galaxies is $n_\mathrm{g} \sim 10^{-3} (h^{-1} \mathrm{Mpc})^{-3}$. This galaxy number density is comparable with what is expected for some upcoming galaxy redshift survey samples, such as DESI's Luminous Red Galaxies which will have $n_\mathrm{g} \approx 0.5 \times 10^{-3} (h^{-1} \mathrm{Mpc})^{-3}$ \citep{2023AJ....165...58Z}.

\section{The ``galaxy infall kinematics'' (GIK) model}
\label{sect:gik}

Our approach to modelling $\xicg$ is built upon the GIK model introduced in \zw. The GIK model describes the velocity distribution of galaxies with respect to nearby clusters, as a function of cluster--galaxy separation. At each cluster--galaxy separation, there are 7 parameters. We will call these 7 parameters the \emph{GIK functions}, because they are functions of the cluster--galaxy separation, $r$, and because this allows us to use the term \emph{GIK parameters} for another set of parameters that we will describe shortly.

\subsection{The GIK functions}

The 7 GIK functions specify the joint probability distribution of radial and tangential velocity components, $P(v_r, v_t | r)$. 
Continuing on from our definitions of $\vect{r}$ and $\vect{v}$ above, we define a unit vector in the separation direction, $\vect{\hat{r}} = \vect{r} / r$.
Then, $v_r = \vect{v} \cdot \vect{\hat{r}}$ is the radial velocity for this cluster--galaxy pair, where negative $v_r$ represents a galaxy moving towards (i.e. falling in to) a cluster. Note that because we use peculiar velocities, the Hubble flow does not contribute to $v_r$; put another way, a cluster--galaxy pair with $v_r = 0$ will have a proper-separation that is growing in time due to the expanding universe. The total tangential velocity is $\vect{v_T} = \vect{v} - v_r \, \vect{\hat{r}}$. For the purpose of the GIK model it is useful to follow \zw in defining the \emph{tangential velocity}, $v_t$, as the projection of $\vect{v_T}$ onto the plane containing the cluster galaxy separation, $\vect{r}$, and the adopted \emph{line of sight}, $\vect{\hat{y}}$. Specifically, $v_t = (\vect{v} \cdot \vect{\hat{y}} - v_r \sin \theta ) / \cos \theta$, where $\theta$ is the angle between $\vect{r}$ and the plane of the sky, i.e. $\sin \theta = \vect{r} \cdot \vect{\hat{y}} / r$.


A diagram of these various vector definitions is provided in Fig.2 of \zw. To aid with intuition, we note that with $v_r$ and $v_t$ defined in this manner, if the distribution of $\vect{v}$ is a zero-mean 3D Gaussian with equal dispersion along three orthogonal axes, then $P(v_r, v_t)$ will be a 2D Gaussian, centred on $v_r = v_t = 0$, and with equal dispersion in the $v_r$ and $v_t$ directions.

\begin{figure}
        \centering
        \includegraphics[width=\columnwidth]{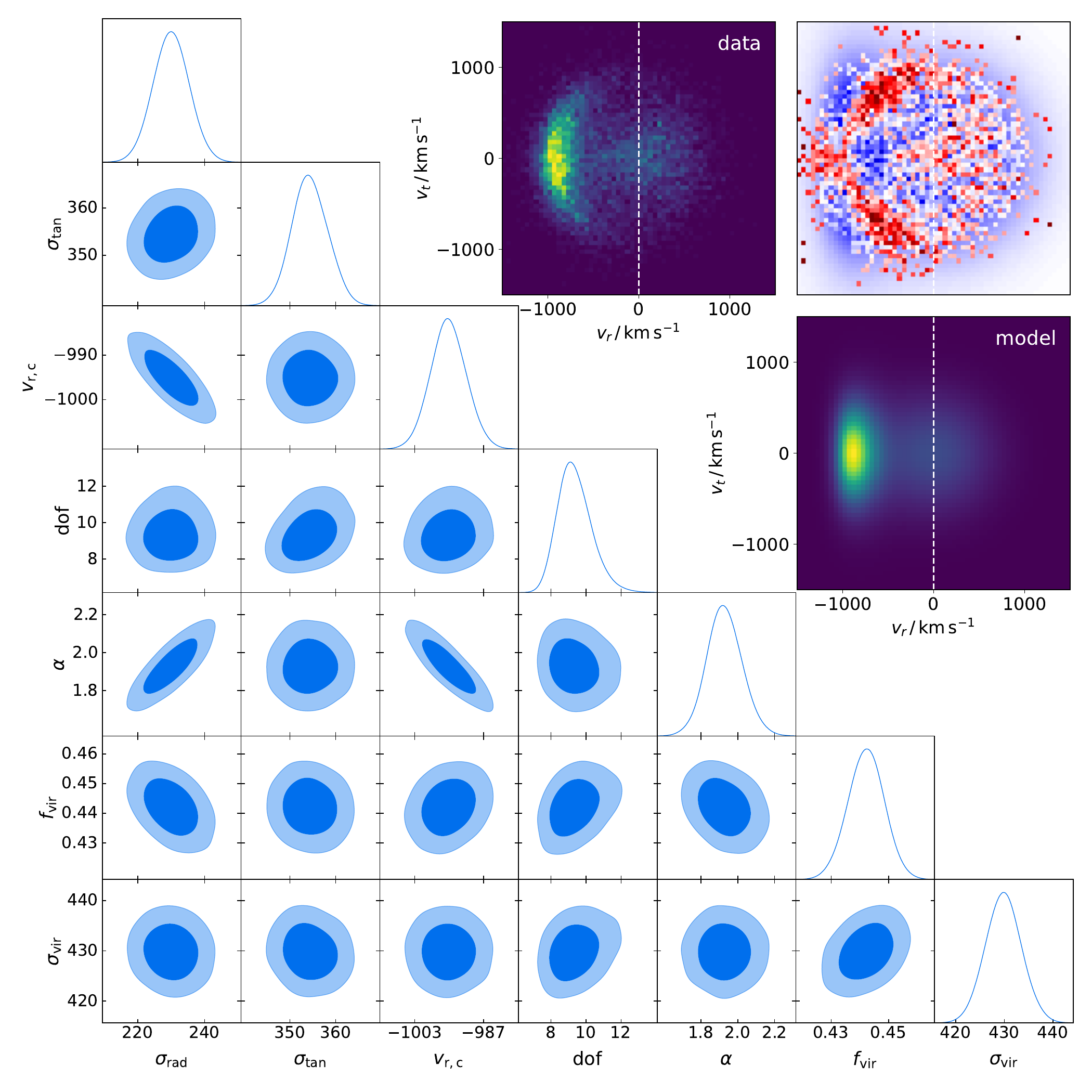}     
	\caption{A corner plot showing the posterior distribution for the GIK functions fit to data generated from a stack of 24 Indra simulations. The cluster sample is all haloes with $14.1 < \log_{10} M_{200} / \hmsun < 14.2$, with galaxies defined as subhaloes with $v_\mathrm{max} > 250 \kms$, and is for cluster--galaxy separations in the range $1 < r / \hmpc < 1.25$. The data and best-fitting model for $P(v_r,v_t)$ are plotted, using the same colour scale in the two panels. The normalised residuals (data minus model, divided by a Gaussian approximation to the Poisson uncertainty) are plotted in the top-right, with red corresponding to positive residuals, and blue negative.}
	\label{fig:Pvrvt_example}
\end{figure}

In Fig.~\ref{fig:Pvrvt_example} we show $P(v_r, v_t)$ for cluster--galaxy separations in the range $1 < r / \hmpc < 1.25$, for a stack of cluster--galaxy pairs from 24 Indra simulations, using the same cluster and galaxy definitions as before. 
This radial shell corresponds to galaxies at distances close to the clusters' virial radii, and an interesting feature of $P(v_r, v_t)$ is that it is bimodal \citep[see also][]{2023MNRAS.521.2464G}. This demonstrates the motivation for the functional form of $P(v_r, v_t)$ in the GIK model, which is the sum of two distinct components: a \emph{virialised} component, making up a fraction $f_\mathrm{vir}$ of galaxies in this radial shell, and an \emph{infalling} component, which makes up the rest. The expectation is that $f_\mathrm{vir}$ is only appreciably non-zero at small radii, approximately within the splashback radii \citep{2015ApJ...810...36M} of the clusters.

The virialised component is modelled with a simple velocity distribution, which is an isotropic Gaussian, with a dispersion along each orthogonal direction of $\sigma_\mathrm{vir}$. The infalling component has a more complex velocity distribution, known as a skewed-$t$ distribution \citep{2009arXiv0911.2342A}. The expression for the infalling component's velocity distribution is
\begin{equation}
  \begin{aligned}
\mathcal{T}(\vect{v}) = 2 \, & t_2(\vect{v} - \vect{\vrc}; \dof, \Sigma) \times  \\ 
& T_1 \left( \frac{\alpha (\vr - \vrc)}{\sigma_\mrm{rad}} \frac{\dof + 2}{Q_v + \dof}; \dof+2 \right),
  \end{aligned}
\label{eq:infalling_Pvrvt}
\end{equation}
where $t_2$ is the density function of a 2D $t$-variate with $\dof$ degrees of freedom,\footnote{Implemented using the \texttt{pdf} method of a \texttt{scipy.stats.multivariate\_t} object.} $T_1$ is the cumulative distribution function for a scalar $t$-variate with $\dof + 2$ degrees of freedom,\footnote{Implemented using the \texttt{cdf} method of a \texttt{scipy.stats.t} object.} $\vect{\vrc} = (\vrc, 0)$ is a typical infall velocity (equal to the mean infall velocity for the case of zero skewness), and $Q_v = (\vect{v} - \vect{\vrc})\transpose \, \Sigma^{-1} \, (\vect{v} - \vect{\vrc})$, where
\begin{equation}
\Sigma = \begin{pmatrix} \sr^2&0\\ 0&\st^2 \end{pmatrix}.
\end{equation}

The basic idea is that $t_2(\vect{v} - \vect{\vrc}; \dof, \Sigma)$ is a 2D $t$-distribution, which looks Gaussian in the limit $\dof \to \infty$, but is leptokurtic (i.e. with fatter tails than a Gaussian with the same variance) for finite $\dof$. This 2D $t$-distribution has variances of $\sr^2$ and $\st^2$ in the radial and tangential directions, respectively. The $t_2$ distribution is symmetric about the typical infall velocity, $\vect{\vrc}$, with the skewness in $\mathcal{T}(\vect{v})$ introduced through multiplying $t_2$ by the cumulative distribution function $T_1$.

The probability density function $\mathrm{d} T_1(x; \dof+2) / \mathrm{d}x$ is symmetric about $x=0$, such that $T_1(x; \dof+2)$ is 0 as $x \to - \infty$, 1 as $x \to \infty$ and $1/2$ at $x=0$. $\alpha$ controls the skewness of the velocity distribution (in the $v_r$ direction), with $\alpha=0$ producing a distribution with no skewness, and positive $\alpha$ producing a distribution with a lengthened tail towards positive $\vr$.

The complete velocity distribution reads
\begin{equation}
P(\vect{v}) = \fvir \, \mathcal{G}(\vect{v}) + (1 - \fvir) \mathcal{T}(\vect{v}),
\label{eq:model_vel_distribution}
\end{equation}
where $\mathcal{G}(\vect{v})$ is a 2D isotropic Gaussian centred on $\vect{v} = 0$. In summary, the 7 GIK functions are:
\begin{itemize}
    \item $\sr$ - the velocity dispersion of the infalling component in the radial direction.
    \item $\st$ - the velocity dispersion of the infalling component in the tangential direction.
    \item $\vrc$ - the typical radial velocity of the infalling component.
    \item $\dof$ - the degrees of freedom associated with the infalling component's 2D $t$ distribution.
    \item $\alpha$ - the skewness in the radial direction of the infalling component's velocity distribution.
    \item $\fvir$ - the fraction of galaxies in the virialised component.
    \item $\sv$ - the 1D velocity dispersion of the virialised component.
\end{itemize}

In Fig.~\ref{fig:Pvrvt_example} we show the posterior distribution for the 7 GIK functions fit to the data shown in the same figure. For a given radial bin, we measure the galaxy velocity distribution by counting galaxies in velocity bins, with $60 \times 60$ equal-sized bins covering $v_r$ and $v_t$ in the range $-1500 \kms \leq v_r, v_t \leq 1500 \kms$. We label the galaxy count in a given pixel $n_{ij}$, where $i$ and $j$ index the pixels along the $v_r$ and $v_t$ directions respectively. To evaluate the likelihood for a given set of GIK function values, we calculate the probability of getting the measured set of $n_{ij}$ assuming that the number of galaxies in a pixel is a Poisson process with an expectation given by the model probability density at the centre of the pixel multiplied by the pixel area and the total number of cluster--galaxy pairs in this radial bin. Specifically, the likelihood is 
\begin{equation}
\mathcal{L} = \prod_{i,j} \lambda_{ij}^{n_{ij}} \exp(- \lambda_{ij}) / n^\mathrm{sim}_{ij} !
\label{eq:Pvrvt_likelihood}
\end{equation}
where $\lambda_{ij} =  P(v_{r,i}, v_{t,j}) (\delta v)^2 N$, with $\delta v = (v_{r,i+1} - v_{r,i}) = (v_{t,j+1} - v_{t,j}) = 50 \kms$ the side-length of a velocity-space pixel, and $N$ the number of cluster--galaxy pairs in this radial shell. Combining this likelihood with flat priors on the 7 GIK functions covering a range considerably broader than the resulting posteriors, we use the MCMC sampler \textsc{emcee} \citep{ForemanMackey:2013io} to find the posterior distribution for the GIK functions. 

We note that in Fig.\ref{fig:Pvrvt_example} the model does a reasonable job of capturing the general features of the measured $P(v_r,v_t)$, however it clearly does not fit the data down to the noise level (associated with a stack of 24 $(1 \, \hgpc)^3$ simulation volumes), with there being clear structure in the map of the residuals. In particular, the infalling component in the data has curvature in the $v_r$--$v_t$ plane, such that galaxies falling in with larger (negative) radial velocities tend to have lower tangential velocities, which is not something that can be captured by the functional form of $\mathcal{T}(\vect{v})$ in eq.~\ref{eq:infalling_Pvrvt}.

\subsection{The GIK parameters}
\label{sect:GIKparams}

\begin{figure*}
        \centering
        \includegraphics[width=\textwidth]{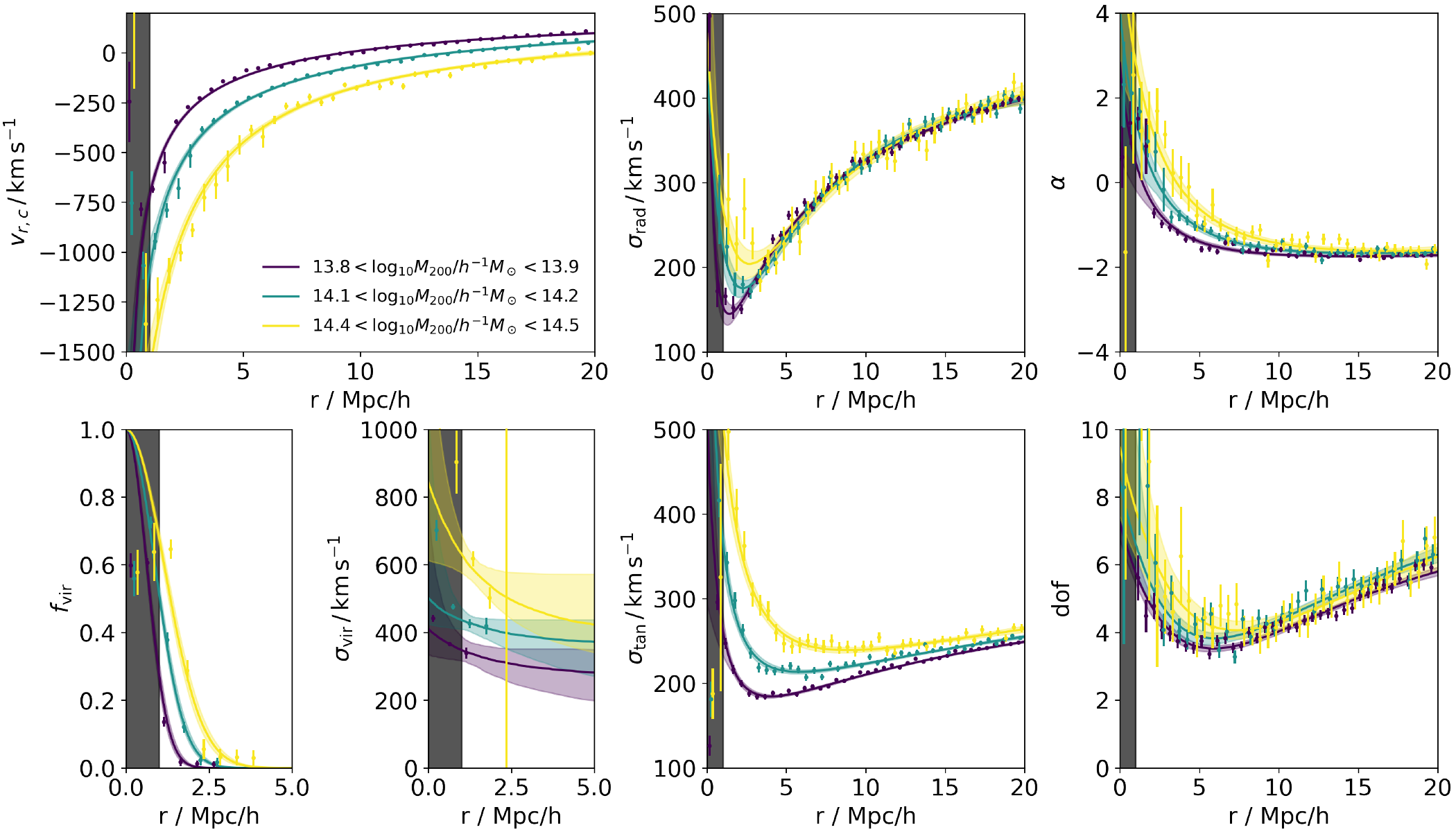}     
	\caption{The variation of the GIK functions with radius. The points with error bars represent the mean $\pm$ one standard deviation of the marginalised posteriors from fitting the GIK functions -- separately at each radius -- to $P(v_r, v_t)$ (for example, as shown in Fig.~\ref{fig:Pvrvt_example}). Generating posterior samples for the GIK parameters (using MCMC) and then calculating the corresponding GIK functions produces the solid lines with shaded regions. For each GIK parameter sample we calculate $\vrc(r)$, $\fvir(r)$, etc. and then the shaded region covers the 2.5th-97.5th percentiles of the GIK function at each radius, with the solid line being the median (50th percentile). The functional forms of the GIK functions are detailed in Appendix~\ref{App:GIK_functions}. We note that $\sv$ is poorly constrained at radii where $\fvir \approx 0$, which leads to the vertical yellow line (an error bar extending off the figure) in the $\sv$ panel.}
	\label{fig:ZW_Fig5}
\end{figure*}

If we perform a fit similar to that shown in Fig.~\ref{fig:Pvrvt_example} for a set of $r$-bins, we can then see how the GIK functions vary with radius. An example of this is shown by the markers in Fig.~\ref{fig:ZW_Fig5}, where now the results are for a single Indra simulation (as opposed to the stack used for Fig.~\ref{fig:Pvrvt_example}, to highlight subtle deficinecies in the model), and we plot the results for three different cluster mass bins, all with the same galaxy definition of any subhalo with $v_\mrm{max} > 250 \kms$.

As well as introducing the form for the velocity distribution in terms of the GIK functions (eq.~\ref{eq:model_vel_distribution}), \zw presented parametric models for how the GIK functions vary with radius. For example, in \zw the functional form for $\sr(r), \st(r), \alpha(r)$ and $\dof(r)$ are all the same:
\begin{equation}
f(r) = q - p \frac{r}{(r+r_i)^\beta}, 
\label{eq:ZW_fr}
\end{equation}
where $q$, $p$, $r_i$ and $\beta$ are free parameters (which take on different values for each of the 4 GIK functions to which this functional form is applied). We note that, written in this way, this particular functional form has the undesirable property that the dimensions of $p$ depend on $\beta$. For this, as well as some other reasons discussed below, we choose to reparameterise this equation. In particular, we write eq.~\ref{eq:ZW_fr} as
\begin{equation}
f(r) = q - h \frac{r }{r_\mrm{min} } \left( \frac{\beta \, r_\mrm{min}}{r + (\beta-1) \, r_\mrm{min} } \right)^\beta,
\label{eq:my_fr}
\end{equation}
where $r_\mrm{min} = r_i/(\beta -1)$ is the radius at which $f(r)$ is minimised, and $h$ can be expressed in terms of $p$, $r_i$ and $\beta$. While this new expression appears more complex than the one it replaces, it leads to free parameters with better characteristics than those of eq.~\ref{eq:ZW_fr}. In particular, $q$ and $h$ have the same dimensions as $f$ (e.g. they are velocities when $f(r)$ is $\st(r)$), and the minimum of $f(r)$ occurs at $r=r_\mrm{min}$, with $f(0) = q$ and $f(r_\mrm{min}) = q-h$. Aside from the reparameterisation being useful for one's intuition ($r_\mrm{min}$ has a clear meaning, while $r_i$ did not), it also decreases the covariance between the model parameters in the fit. As an example, the location of the minimum of $\sr(r)$ in Fig.~\ref{fig:ZW_Fig5} is clearly well constrained by the data, such that we get a tight constraint on $r_\mrm{min}$, whereas this would correspond to some degenerate set of solutions in terms of $r_i$ and $\beta$.

For similar reasons, we reparameterised the functional forms for the majority of GIK functions from those that appear in \zw. We describe this in Appendix~\ref{App:GIK_functions}. In total we end up with 24 parameters that describe the variation of the 7 GIK functions with radius; we call these 24 parameters the \emph{GIK parameters}.

Note that if one simply fits the 7 GIK functions to $P(v_r,v_t)$ at large radii ($\gtrsim 10 \, \hmpc$) the values of $\fvir$ tend to be fairly large ($\gtrsim 0.3$). Physically, these clearly do not correspond to galaxies ``virialised'' within the cluster, but instead result from the fact that sufficiently far from the cluster the infalling component has a small mean radial velocity, such that it can be approximately described by the $\mathcal{G}(\vect{v})$ function centred on $v_r = 0$. For the purpose of plotting the GIK function posteriors (points with error bars) in Fig.~\ref{fig:ZW_Fig5} we circumvent this problem by first fitting the full 7 parameter model for $P(v_r,v_t)$ in our radial bins and finding the radius at which $\fvir$ is smallest, which we label $r_f$. We then fit just a 5 parameter model (with $\fvir = 0$) to all radial bins with $r > r_f$.



\subsubsection{Fitting for the GIK parameters}
\label{sect:direct_GIKp_fit}

While the process visualised in Fig.~\ref{fig:ZW_Fig5} -- of measuring the 7 GIK functions in different $r$-bins -- is useful conceptually, in practice we fit the 24 GIK parameters directly to the $P(v_r, v_t | r)$ data, simultaneously fitting to all radial bins. As a default we use 40 $r$-bins, with the bin edges uniformly spaced from 0 to 20 $\hmpc$. The likelihood we use is the product over the different radial bins of the likelihood defined in eq.~\ref{eq:Pvrvt_likelihood}, with the model $P(v_r, v_t)$ at all radii calculated directly from a common set of GIK parameters. 

This direct fitting procedure has a number of advantages over a two-step process of finding the GIK functions at different radii and then fitting the GIK parameters to these (which would amount to fitting the curves described by the GIK parameters to the points with error bars in Fig.~\ref{fig:ZW_Fig5}). In particular, we can naturally account for the covariance between the different GIK parameters, even those relating to different GIK functions; we avoid the need to fix $\fvir = 0$ at large $r$; we avoid the need to define priors on the GIK functions, instead only requiring priors on the GIK parameters; and we also avoid taking the (potentially) non-Gaussian GIK function posteriors, and treating them as Gaussian for the purpose of then fitting the GIK parameters.

\subsubsection{Fitting the GIK parameters without binning}

An alternative to the likelihood defined in eq.~\eqref{eq:Pvrvt_likelihood} is just a product over all cluster-galaxy pairs of the GIK model's probability density for each pair's velocity,
\begin{equation}
\mathcal{L} = \prod_{i} P(\vect{v}_i | r_i, \vect{g}),
\label{eq:likelihood_no_binning}
\end{equation}
where $P(\vect{v}_i)$ is defined in eq.~\eqref{eq:model_vel_distribution} and depends on the set of GIK parameters ($\vect{g}$) and the separation of cluster-galaxy pair $i$ ($r_i$). The fact that we could define a likelihood that did not require either radial or velocity-space binning only occurred to us late in this project, and so the work throughout this paper uses the likelihood from eq.~\eqref{eq:Pvrvt_likelihood}. Nevertheless, we mention the alternative in case it is useful for other people trying to implement a similar modelling method. We tested that the likelihoods in equations \eqref{eq:Pvrvt_likelihood} and \eqref{eq:likelihood_no_binning} produced consistent measurements of the GIK parameters for a fiducial Indra simulation volume, finding that the differences in best-fitting parameters were negligible, with the no-binning likelihood leading to slightly tighter constraints on some GIK parameters, presumably due to information loss when binning. This test is described further in Appendix~\ref{App:no_binning_likelihood}.

\subsection{Convergence of the GIK parameters}
\label{sect:TNG_convergence}

As described later in this paper, we will ultimately use $N$-body simulations run with different cosmologies to predict how the GIK parameters depend upon cosmology. It is therefore important to understand how robust the GIK parameters measured from an $N$-body simulation are. To address this, we use publicly available simulations from the IllustrisTNG project \citep{2019ComAC...6....2N}. In particular, we use variants of the TNG300 simulations, to assess how robust the GIK parameters are to changing simulation resolution, as well as whether the GIK parameters change between dark matter only simulations and simulations including baryonic physics.

For this section, we use the TNG300-1 simulation (which we call ``TNG300'') as our fiducial simulation. This simulation includes both dark matter and baryons, with dark matter particle masses of $5.9 \times 10^7 \msun$ and initial gas masses of $1.1 \times 10^7 \msun$. The TNG300-3 simulation (which we call ``TNG300 low-res'') employs the same physics, but with 64 times fewer particles than TNG300-1, leading to DM masses of $3.8 \times 10^9 \msun$ and gas masses of $7.0 \times 10^8 \msun$. Finally, TNG300-1-Dark (which we call ``TNG300 DMO'') is a dark matter only version of TNG300-1, with a dark matter particle mass of $7.0 \times 10^7 \msun$.

\begin{figure*}
        \centering
        \includegraphics[width=\textwidth]{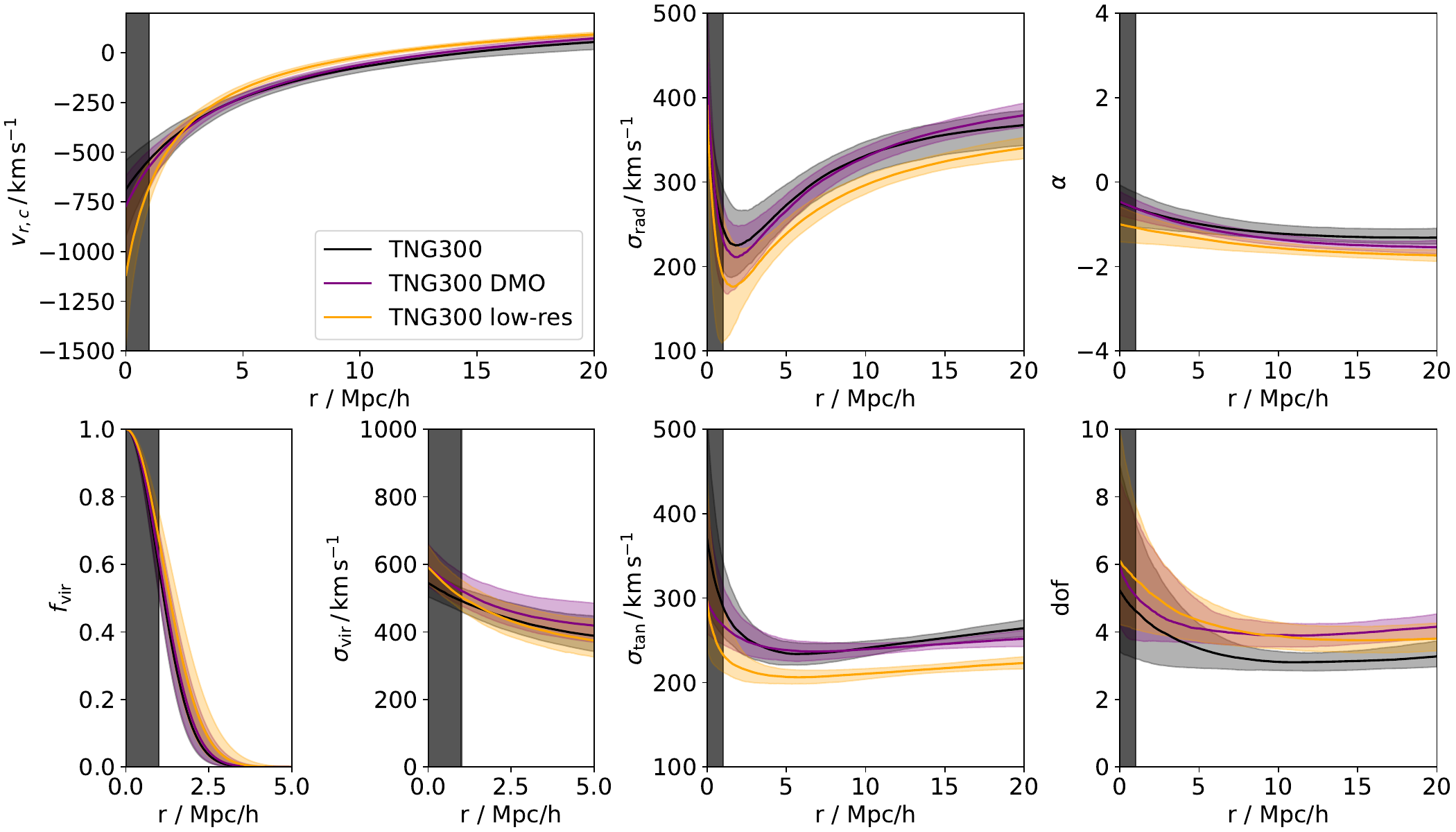}     
	\caption{The GIK functions measured from three different TNG300 simulations. The inclusion of baryonic physics or not seems to have little impact on the GIK functions, while the lower-resolution simulation has slight differences, including a suppression of the radial and tangential velocity dispersion of the infalling componenet.}
	\label{fig:TNG_ZW_Fig5}
\end{figure*}

For each of these simulations we used the method described in Section~\ref{sect:direct_GIKp_fit} to fit for the GIK parameters. We ran an MCMC to generate samples from the posterior distribution for the GIK parameters. By calculating the corresponding GIK functions for each MCMC sample, we generate draws of the GIK functions from this posterior. These are plotted in Fig.~\ref{fig:TNG_ZW_Fig5}, where the solid lines show the median values of the GIK functions, while the shaded regions cover the 16th to 84th percentiles of the GIK function posteriors.  

In Fig.~\ref{fig:TNG_ZW_Fig5} the differences between the fiducial and dark matter only simulations are negligible. The lower-resolution simulation does appear to have somewhat different GIK functions, although the differences (except for $\st$ at large radii) are typically less than the uncertainty, where the size of the uncertainty comes from the relatively small box size of the TNG300 simulations.

\begin{figure}
        \centering
        \includegraphics[width=\columnwidth]{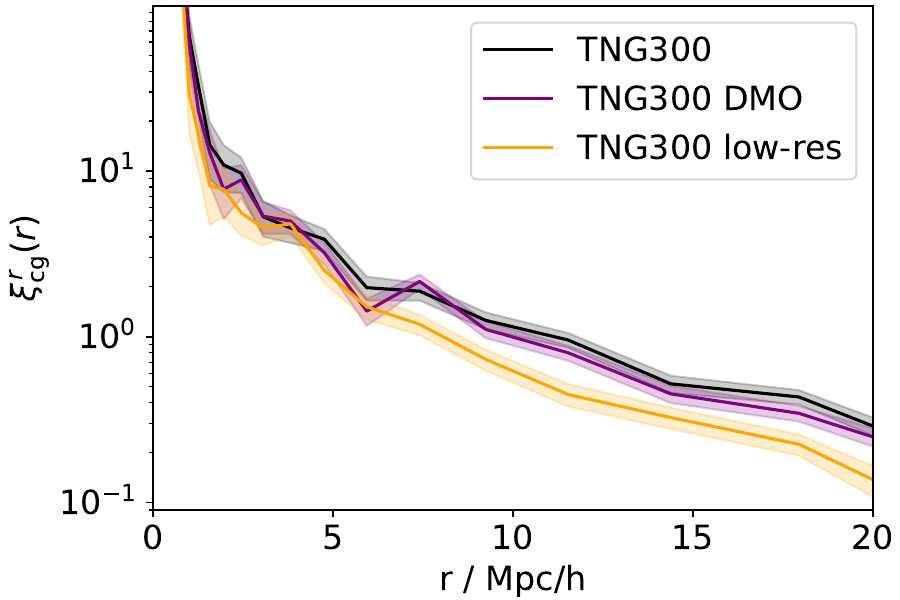}     
	\caption{The real space clustering measured from three different TNG300 simulations. The inclusion of baryonic physics has little impact, while the lower-resolution simulation has suppressed clustering at large separations. The correlation function was measured by counting cluster-galaxy pairs in bins of separation and comparing with the expectation if clusters and galaxies were randomly distributed. The shaded region is an indication of the uncertainty on the correlation function measurement, from assuming that the pair counts follow Poisson statistics.}
	\label{fig:TNG_clustering}
\end{figure}

In Fig.~\ref{fig:TNG_clustering} we plot the real space clustering from these three different TNG300 simulations. As described in detail in Section~\ref{sect:calc_xicg_with_GIK}, the combination of the real space clustering and the GIK functions (or some other description of the pairwise velocity distribution) are all that is required to calculate $\xicg$. We see that, as was the case for the GIK functions, TNG300 and TNG300 DMO produce similar results, with TNG300 low-res being more discrepant.

We note that the GIK parameter differences between dark matter only vs hydro simulations, or high-resolution vs low-resolution, are significantly less important than the differences in real space clustering between those same simulations for evaluating a model $\xicg$. This is quantified in Appendix~\ref{App:TNG_convergence}, and suggests that relying on simulations to predict the GIK parameters -- while using some other method for the real space clustering -- is a sensible approach. This is what we do throughout the rest of the paper, where (because we are dealing with simulations) we know the real space clustering and so use the true real space clustering in our models for $\xicg$. With real observational data, we would instead need a model for the real space clustering, or to obtain it from observations of the projected clustering. These possibilities are discussed in Section~\ref{sect:real_space_clustering}.

\section{Calculating $\xicg$ with the GIK model}
\label{sect:calc_xicg_with_GIK}

In the previous section we described the GIK model from \zw, and presented a method to fit the GIK parameters to the positions and velocities of clusters and galaxies extracted from a simulation. We will now show how, with the GIK parameters determined, we can predict the clustering in redshift-space given the real-space clustering.

We begin with the expression for the redshift-space clustering in terms of the real space clustering and the probability distribution for $\rpi - y$ (i.e. for how a cluster--galaxy pair's separation is shifted when moving from real to redshift space). This usually goes by the name of the \emph{streaming model}, and can be written as \citep{1980lssu.book.....P, 2004PhRvD..70h3007S, 2018MNRAS.479.2256K}
\begin{equation}
1 + \xicg(\rp, \rpi) = \int_{-\infty}^\infty \left[ 1 + \xi_\mathrm{cg}^r(r) \right] p(\rpi-y | \rp, y) \, \mathrm{d}y,
\label{eq:streaming_model}
\end{equation}
where $r = \sqrt{\rp^2 + y^2}$, and $p(\rpi-y | \rp, y) = p(v_y | \rp, y) H(z)/(1+z)$ is the probability density associated with a given pairwise velocity along the line-of-sight, where the factor of $H(z)/(1+z)$ comes from $\mathrm{d}\rpi / \mathrm{d}v_y$.

The distribution of radial and tangential velocities at a given separation, $p(v_r, v_t | r)$, is what is specified by the GIK model. The conversion from $(v_r, v_t)$ to $v_y$ depends on the angle between the line-of-sight and the cluster--galaxy separation, and so depends on $\rp$ and $y$. Mathematically, we have $v_y = v_r \sin \theta + v_t \cos \theta$, where $\theta = \tan^{-1} y/\rp$ is the angle between the plane-of-the-sky and the cluster--galaxy separation. This leads to $p(v_y | \rp, y)$ being expressible as an integral:
\begin{equation}
p(v_y | \rp, y) = \int_{-\infty}^\infty p\left( v_r, v_t = \frac{v_y - v_r \sin \theta}{\cos \theta} \, | \, r \right) \, \frac{\mathrm{d}v_r}{\cos \theta}.
\label{eq:vy_from_vr_vt}
\end{equation}
Combining equations~\ref{eq:streaming_model} and \ref{eq:vy_from_vr_vt}, and integrating only out to finite limits, we have 
\begin{multline}
1 + \xicg(\rp, \rpi) \approx \frac{H(z)}{1+z} \times \\ \int_{y_\mrm{min}}^{y_\mrm{max}} \frac{1 + \xi_\mathrm{cg}^r(r)}{\cos \theta} \int_{v_\mrm{min}}^{v_\mrm{max}} p\left( v_r, v_t | r \right) \, \mathrm{d}v_r \, \mathrm{d}y,
\label{eq:xicg_double_integral}
\end{multline}
where again $v_t = (v_y - v_r \sin \theta)/ \cos \theta$ and $v_y = H(z) (\rpi - y) / (1+z)$. In order for eq.~\ref{eq:xicg_double_integral} to be exact, the integration limits should cover an infinite range of $y$ and $v_r$. However, for the purpose of calculating $\xicg(\rp, \rpi)$ for $\rp, \rpi < 20 \, \hmpc$ (which we do here) we found that $y_\mrm{max} = - y_\mrm{min} = 40 \, \hmpc$ and $v_\mrm{max} = - v_\mrm{min} = 2500 \, \kms$ were sufficient. 
We note that the reason for using finite limits was that to speed up the calculation, we did not use a generic numerical integration routine (which could potentially handle infinite limits), instead using fixed $v_r$, $v_t$ and $y$ grids and evaluating equation~\ref{eq:xicg_double_integral} for a large number of $\rpi$ (but fixed value of $\rp$) simultaneously, by summing up the integrand evaluated over the grids. This process is much faster because multiple evaluations of $p(v_r, v_t | r)$ with a particular $r$ can be made simultaneously.\footnote{Due to how the \texttt{pdf} method of the \texttt{scipy.stats.multivariate\_t} class is written, it is much faster to evaluate $p(v_r, v_t | r)$ with thousands of different $(v_r, v_t)$ combinations simultaneously, than to make thousands of separate calls to $p(v_r, v_t | r)$.} We use 1000 equally spaced $v$-bins (with $5 \, \kms$ spacing) and an adaptive number of $y$-bins, that we double until successive iterations agree on all values of $\xicg(\rp, \rpi)$ to within a specified tolerance, which by default we set to be a fractional agreement of better than 1\%.


\section{Building an emulator for the GIK parameters}
\label{sect:GIK_param_emulator}

One might hope that the GIK functions and their dependence on cosmological parameters can be predicted from theory (i.e. ``with pen and paper''). However, even predicting the mean infall velocity onto clusters (closely related to $\vrc$) from first principles is not easy, with simple spherical infall models failing to correctly predict the results of $N$-body simulations for the mean infall velocity of dark matter particles as a function of their distance from the centres of clusters \citep[e.g. Fig.~3 of][]{1986ApJ...308..499V}. However, the fact that the GIK functions at fixed cosmology seem not to be too sensitive to the physics employed within the simulation, or to the numerical resolution at which the simulation was run (see Fig.~\ref{fig:TNG_ZW_Fig5}), suggests that the GIK functions can be inferred from $N$-body simulations in a reasonably reliable manner.

This motivates using suites of simulations (run with different cosmological parameters) to build an emulator that can predict how the GIK parameters depend upon cosmology. Combining such an emulator with knowledge of the real-space clustering of galaxies and clusters, $\xi_\mathrm{cg}^r(r)$, allows us to predict the redshift-space clustering as a function of cosmology, and therefore to fit the cosmological parameters to an observed $\xicg$ data vector.

\subsection{Comparison with directly emulating the $\xicg$ data vector}
\label{sect:pros_cons_vs_direct_xicg}

There have been a number of recent attempts to model redshift-space galaxy clustering on small scales using emulators trained on $N$-body simulations \citep{2019ApJ...874...95Z, 2022MNRAS.515..871Y, 2023MNRAS.520.5373L, 2023arXiv230212379K}. Aside from the fact that these have focused on galaxy-galaxy rather than cluster-galaxy correlations, our approach is also conceptually quite different. In particular, these other works directly measure observables (e.g. $\xi_\mathrm{gg}^s$ on a grid of $r_p$, $r_\pi$, or multipoles of $\xi_\mathrm{gg}^s$) from simulations, and then build emulators for these observables as a function of the cosmological parameters,\footnote{For brevity, we refer only to the cosmological parameters, but the emulators also include parameters for the so-called \emph{galaxy-halo connection} \citep{2018ARA&A..56..435W}, which describes how galaxies populate dark matter halos.} while we break up the calculation into two distinct parts (obtaining the real-space clustering, and then modelling the RSD effects on this), with an emulator used to do the second part.\footnote{We note that this approach is similar to that described in \citet{2023MNRAS.523.3219C}, where they present an emulator for real space clustering, that they intend to combine with an emulator for a parameterised velocity distribution similar to the ``infalling'' part of the GIK model (equation~\ref{eq:infalling_Pvrvt}), described in \citet{2020MNRAS.498.1175C}.}

\subsubsection{Reducing noise with the GIK model}
\label{sect:reducing_noise}

One considerable benefit of our approach is that assuming a functional form for the pairwise velocity distribution of cluster-galaxy pairs acts to reduce the effects of noise in each simulation used to build the emulator. This enables the use of smaller simulations than would be required if one simply ``observes'' each simulation in a similar manner to how the real observations will be made.

To demonstrate the benefit of this quantitatively, we begin by defining $\xi^{s,k}_\mrm{sim}$, the $\xicg$ data vector calculated directly from the $k$-th Indra simulation (with an example plotted in the right-hand panel of Fig.~\ref{fig:example_xi_cg}). We additionally define the data vector calculated by combining the real space clustering from the $k$-th Indra simulation, $\xi^{r,k}(r)$, with the maximum-likelihood GIK parameters from the $k$-th Indra simulation, as $\xi^{s,k}_\mrm{GIK}$. 
We also define the Indra-derived covariance matrix for the elements of $\xicg$ as $C_\mrm{sim}$. This is calculated from the set of $\xi^{s,k}_\mrm{sim}$ (with some Jackknifing, described in Section~\ref{sect:cov}). The inverse of $C_\mrm{sim}$ is the precision matrix, $\Psi_\mrm{sim} = C_\mrm{sim}^{-1}$, with the elements of $\Psi_\mrm{sim}$ that relate to the first 20 elements of our flattened data vector (which is all those with $r_p = 0.5 \, \hmpc$) set to zero. This means that we give no weight to the lowest-$r_p$ column of the $\xicg$ maps, because the parametric GIK functions do not agree with the measured GIK functions on scales below $1 \, \hmpc$ (see Fig.~\ref{fig:ZW_Fig5}). In addition, we define the values along the diagonal of $C_\mrm{sim}$ to be $\sigma_\xi^2$.

\begin{figure}
        \centering
        \includegraphics[width=\columnwidth]{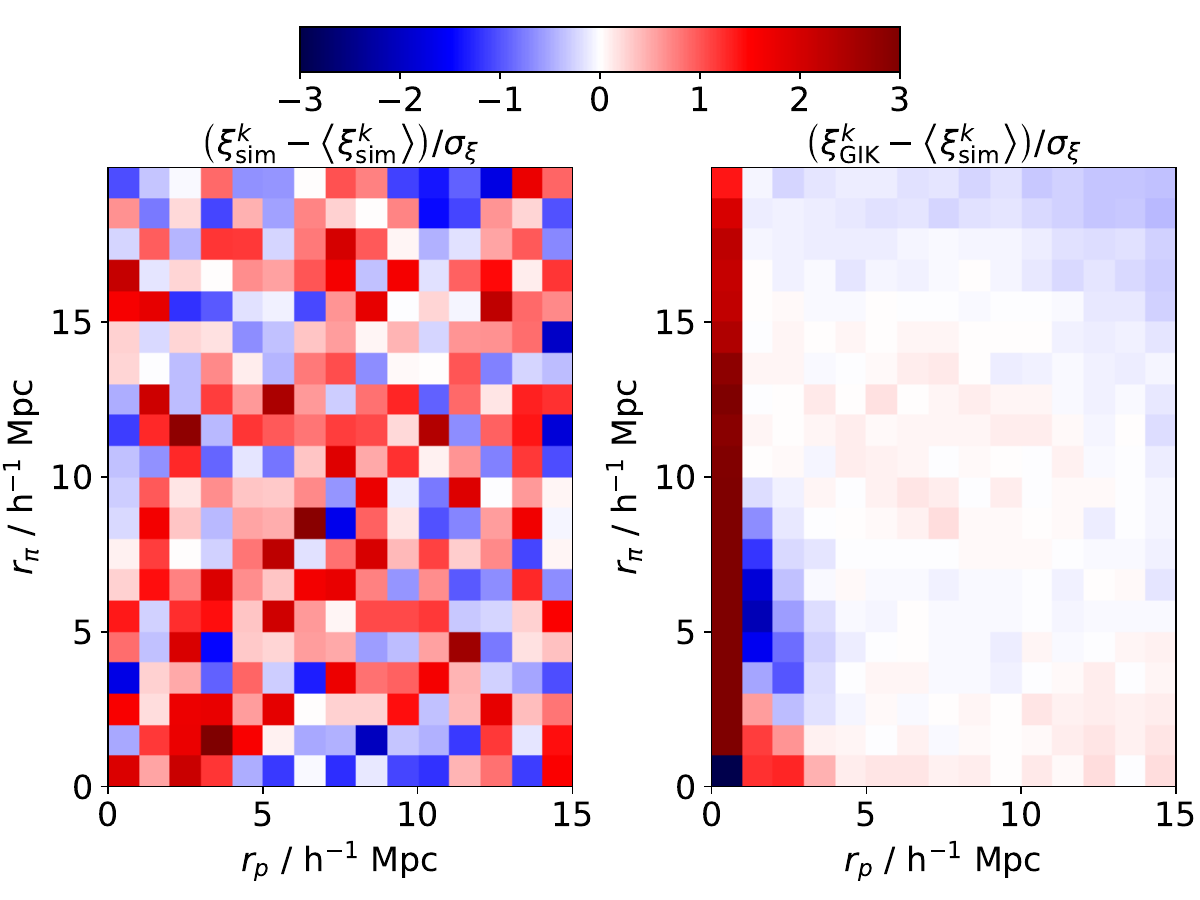}     
	\caption{Maps of the normalised residuals, comparing $\xicg$ from one example Indra simulation, to the average over all Indra simulations. In the left panel $\xicg$ from the single simulation is calculated directly from the cluster and galaxy positions in redshift space in the simulation, while in the right panel the GIK parameters are fit to the simulation data, which is then combined with the real-space clustering to produce $\xicg$. This second procedure dramatically decreases the amount of noise on $\xicg$ measured from a single simulation, though model misspecification leads to systematic residuals at low-$r_p$.}
	\label{fig:GIK_vs_sim}
\end{figure}

With these definitions, we can assess how precisely we can use a single Indra simulation to estimate the true $\xicg$ for the Indra cosmology, which we assume is well approximated by $\left< \xi^{s,k}_\mrm{sim} \right>$, where the expectation value is an average over all of the Indra simulations. In particular, we define 
\begin{equation}
\begin{split}
\chi^{2,k}_\mrm{sim} & = \left(\xi^{s,k}_\mrm{sim} - \left< \xi^{s,k}_\mrm{sim} \right>\right)\transpose \Psi_\mrm{sim} \left(\xi^{s,k}_\mrm{sim} - \left< \xi^{s,k}_\mrm{sim} \right>\right) \\
\chi^{2,k}_\mrm{GIK} & = \left(\xi^{s,k}_\mrm{GIK} - \left< \xi^{s,k}_\mrm{sim} \right>\right)\transpose \Psi_\mrm{sim} \left(\xi^{s,k}_\mrm{GIK} - \left< \xi^{s,k}_\mrm{sim} \right>\right).
\end{split}
\end{equation}
Calculating these $\chi^2$ values for each Indra simulation, we find that $\left< \chi^{2,k}_\mrm{sim} \right> = 301$, while $\left< \chi^{2,k}_\mrm{GIK} \right> = 27$.
This means that employing the GIK model, we can calculate a model $\xicg$ data vector from a single Indra simulation that is much closer (by a factor of $\sim 10$ in $\chi^2$) to the average $\xicg$ directly measured from all Indra simulations, than $\xicg$ directly measured from any one Indra simulation is to the Indra average. 

For some intuition as to how this works, in Fig.~\ref{fig:GIK_vs_sim} we plot the \emph{normalised residuals} for one example Indra simulation, which we define as the difference between a $\xi^{s,k}$ predicted from one simulation and the average $\xi^{s,k}_\mathrm{sim}$ over all Indra simulations, divided by the simulation-to-simulations noise ($\sigma_\xi$). In the left-hand panel we do this for a case where the one-simulation prediction is measured directly from the simulation ($\xi^{s,k}_\mrm{sim}$), while in the right-hand panel the one-simulation prediction uses the GIK model ($\xi^{s,k}_\mrm{GIK}$). As expected, the map on the left consists of many fluctuations of order $1 \sigma_\xi$, while employing the GIK model smooths over this noise in a physically motivated way.

\subsubsection{Model misspecification error}
\label{sect:model_misspec}

The main drawback of our approach is also highlighted in Fig.~\ref{fig:GIK_vs_sim}, and is the possibility for model misspecification error. In particular, if the GIK model does not adequately describe the pairwise velocity distribution of cluster-galaxy pairs, then using this model to map from real-space to redshift-space clustering can introduce systematic errors. The residuals seen at $\rp \lesssim 3 \, \hmpc$ in the right-hand panel of Fig.~\ref{fig:GIK_vs_sim} demonstrate some deficiency of the GIK model, and could arise for two primary reasons. One is if the 7 GIK functions do not adequately describe $P(v_r, v_t | r)$ at fixed radius, while the other is if the 24 GIK parameters do not adequately describe the radial dependence of the GIK functions.

We have already observed that with a sufficiently large simulated volume, deviations between $P(v_r, v_t | r)$ from simulations, and the 7-parameter velocity distribution function at the heart of the GIK model (eq.~\ref{eq:model_vel_distribution}) can be detected, as demonstrated in the residuals panel of Fig.~\ref{fig:Pvrvt_example}. In Appendix~\ref{App:model_misspec} we investigate the relative importance of these two effects, finding that the GIK parameters' inability to perfectly describe the GIK functions explains about half (in a $\chi^2$ sense) of the residuals, with the remaining half coming from the 7-parameter model for $P(v_r, v_t | r)$.

Realistic surveys are likely to have significantly greater uncertainties on these small scales than implied by $\sigma_\xi$, because spectroscopic surveys face significant observational challenges when dealing with close pairs of galaxies \citep[e.g.][]{2017MNRAS.472.1106B}. This suggests that the model misspecification errors that appear highly salient in Fig.~\ref{fig:GIK_vs_sim} may not matter so much in a real survey. Also, the model misspecification errors are confined to a restricted region of the $r_p$-$r_\pi$ plane, so an analysis using the GIK model could cut out the unreliable regions without much difficulty. Depending on the accuracy requirements for the model $\xicg$ when applied to real observational data, it may be that alternative functional forms that can better describe the velocity distribution are required, which we leave to future work.

\subsubsection{Separating out real space clustering from RSD effects}

Our approach of building a model for the mapping from real space to redshift space clustering and then combining this with some model/measurement of the real space clustering to get $\xicg$ is conceptually appealing, because physically the real space clustering and the RSD are separate processes. This makes our model more interpretable than an emulator that directly predicts $\xicg$, and aids when trying to understand intuitively the effects of different cosmological parameters/models on $\xicg$.

Additionally, there are observables that depend on the real space clustering (such as cosmic shear \citep{2015RPPh...78h6901K}), as opposed to the redshift space clustering. Our approach means that when doing a joint analysis of different cosmological observables, consistent real space clustering can be used in the modelling of the different observables, whereas when directly emulating $\xicg$, the underlying real-space clustering may be inconsistent with that used elsewhere in the analysis.

\subsubsection{Simpler cosmological parameter dependence of emulated quantities}

While we do not attempt to quantify it here, another potential advantage of emulating the GIK functions (rather than $\xicg$ directly) is that they are simpler physical quantities than $\xicg$ in a particular $(r_p, r_\pi)$ pixel. This means that one would expect the GIK parameters describing the GIK functions to vary in a more systematic way with variations to the cosmological parameters than the redshift-space correlation function does. For example, if increasing some cosmological parameter leads to an increase in the infall velocities, $|\vrc|$, this would lead to systematic shifts in the relevant GIK parameters. But in terms of values of $\xicg$ at particular $(\rp, \rpi)$ this could behave in a non-monotonic manner, with (for example) ``Kaiser squashing'' enhancing the clustering, before ``fingers of god'' decrease it. While Gaussian processes are flexible enough to model more complex functions, emulation of simpler functions can be done with less thought given to the Gaussian process kernels, and should require fewer training data locations to produce a robust emulator.

\subsection{The FORGE simulations}

One of the major goals of modern cosmology is to understand what drives the accelerated expansion of the Universe at late times \citep{2013PhR...530...87W}, with modified gravity theories being a popular alternative to GR plus a cosmological constant. $f(R)$ gravity \citep[first proposed in][]{1970MNRAS.150....1B} is one popular modified gravity theory, which we investigate here as an example of how one would use the GIK model to constrain the parameters of a particular theory of gravity. We note that of the many modified gravity theories that have been proposed, we do not hold $f(R)$ in any special regard, but choose to focus on $f(R)$ in the first instance due to the fact that the GIK parameters have already been shown to differ in $f(R)$ and GR \citep{2014MNRAS.445.1885Z}, and due to the existence of FORGE \citep{2022MNRAS.515.4161A}, a suite of cosmological simulations run with different $f(R)$ models.



The FORGE simulations were run with 50 different combinations of cosmological parameters in Hu-Sawicki $f(R)$ gravity \citep{2007PhRvD..76f4004H}, including various values for $\fr0$,\footnote{$\fr0$ is the background value of the scalar degree of freedom at $z=0$ and controls the potential depth threshold at which the chameleon screening becomes active and GR-like forces are recovered. Larger values of $|\fr0|$ correspond to larger deviations away from GR.} as well as $\Om$, $\sigma_8$ and $h$. For each simulated cosmology, there are both high-resolution and low-resolution simulations. We only use the high-resolution simulations, which are of cubic volumes with a comoving side length of $500 \, \hmpc$. The particle mass used in these simulations is $9.1 \times 10^9 \, \hmsun$, with a gravitational softening length of $15 \, \hkpc$. We do not use the large volume, low-resolution, simulations because the comparatively poor mass resolution ($m_\mathrm{DM} = 1.5 \times 10^{12} \, \hmsun$) means that the (sub-)structures capable of hosting galaxies of interest to us are not all resolved. As such, we would need a different method of populating our simulations with galaxies in order to calculate the GIK parameters.\footnote{These larger volume simulations could, for example, be used in conjunction with a Halo Occupation Distribution (HOD) model \citep[e.g.][]{2000MNRAS.318.1144P}, although doing this in the context of extracting the GIK parameters -- particularly those important for the velocity distribution at small cluster-galaxy separations -- would be somewhat circular, as HOD galaxies are typically assigned velocities drawn from simple Gaussian velocity distributions \citep[e.g][]{2015MNRAS.446..578G}.}

The FORGE cosmologies are primarily arranged in a latin hyper-cube, with 49 of the 50 simulations uniformly covering the following parameter ranges:
\begin{itemize}
    \item $\Omega_\mathrm{m}$ from 0.11 to 0.55
    \item $S_8 \equiv \sigma_8 \times (\Omega_\mathrm{m} / 0.3)^{0.5}$ from 0.60 to 0.90
    \item $h$ from 0.60 to 0.82
    \item $\log_{10} |\fr0|$ from -6.2 to -4.5
\end{itemize}
the remaining simulation is a $\Lambda$CDM one, i.e. it has $|\fr0|=0$. Because the majority of the FORGE simulations are evenly distributed in $\log_{10} |\fr0|$ rather than $|\fr0|$, it makes sense to treat $\log_{10} |\fr0|$ as the $f(R)$ parameter in our emulator. However, this then leads to the value for the $\Lambda$CDM simulation being undefined. To avoid this, we instead use $\log_{10} (|\fr0| + 10^{-6})$ as the $f(R)$ parameter in our emulator.


For each of the 50 FORGE simulations, we calculate a posterior distribution for the GIK parameters following the method described in Section~\ref{sect:direct_GIKp_fit}. The priors on the GIK parameters are given in Appendix~\ref{App:GIK_functions}. We continue to define galaxies as halos/subhalos with $v_\mathrm{max} > 250 \kms$, and we evaluate the GIK parameters for 7 different cluster mass bins for each cosmology. Each mass bin spans 0.1 dex in mass, covering the range $13.8 < \log_{10} M_{200} / \hmsun <  14.5$. For each cluster mass bin, we consider the cluster mass associated with that bin to be the geometric mean of the mass of the upper and lower edge of the bin, such that the bin containing halos with $14.1 < \log_{10} M_{200} / \hmsun < 14.2$ is considered to correspond to clusters of mass $10^{14.15} \, \hmsun$.



\subsection{Emulation with Gaussian Processes}

The aim of our emulator is to predict the cluster-galaxy pairwise velocity distribution (and its dependence on cluster-galaxy separation) as a function of the cosmological parameters plus the cluster mass (going forward we will refer to cluster mass as an additional `cosmological parameter' for brevity, such that we have 350 different simulated `cosmologies'). Within the GIK model, the velocity distribution at all radii is specified by the 24 GIK parameters, and so we emulate the velocity distribution by separately emulating each of the 24 GIK parameters.

For the purpose of emulating each GIK parameter, $g_i$, we use a Gaussian process, using the Python package \george \citep{2015ITPAM..38..252A}. A Gaussian process is a prior over possible functions, which, when conditioned on a set of (potentially uncertain) observations of the function (which we will call the \emph{training data}), can produce a posterior distribution for the target function. In our case, the functions for which we would like to infer the posteriors are the $g_i (\vect{\theta})$, with $i$ (ranging from 1 to 24) being the index for a particular GIK parameter, and $\vect{\theta}$ being a vector of cosmological parameters:
\begin{equation}
\vect{\theta} = \{ \Om, \, S_8,\, h,\, \log_{10} (|\fr0|+10^{-6}),\, \log_{10} M_{200} / \hmsun \}.
\label{eq:cosmo_theta}
\end{equation}

A detailed description of Gaussian processes can be found in \citet{2006gpml.book.....R}.\footnote{For intuition on how and why Gaussian processes work, we recommend an excellent lecture by Richard Turner, available on \href{https://www.youtube.com/watch?v=92-98SYOdlY}{YouTube}.} In general, when combined with training data (in our case a mean and Gaussian uncertainty on each of the $g_i (\vect{\theta}_j)$, where the set of $\vect{\theta}_j$ are the 350 input cosmologies at which we have fit for the GIK parameters) a Gaussian process can make a prediction for how the output of a function depends on the input parameters (i.e $g_i (\vect{\theta})$ for arbitrary $\vect{\theta}$).


Specifying a Gaussian process requires that we specify an a priori most-likely \emph{mean function}, $m(\vect{\theta})$, as well as a \emph{covariance function}, often referred to as the \emph{kernel}. This kernel, $K(\vect{\theta}_1, \vect{\theta}_2)$, describes how correlated we expect $g_i (\vect{\theta}_1)$ and $g_i (\vect{\theta}_2)$ to be, as a function of the separation between $\vect{\theta}_1$ and $\vect{\theta}_2$. The kernel is used to specify the notion that we expect $g_i (\vect{\theta})$ to vary smoothly with the cosmological parameters, such that two similar values of $\vect{\theta}$ will have similar values for $g_i (\vect{\theta})$. We use a squared-exponential kernel, which can be written as
\begin{equation}
K(\vect{\theta}_1, \vect{\theta}_2) = \sigma_K^2 \exp \left( - \frac{1}{2} (\vect{\theta}_1 - \vect{\theta}_2)\transpose \vect{\Sigma}_K^{-1} (\vect{\theta}_1 - \vect{\theta}_2) \right) .
\label{eq:kernel_function}
\end{equation}
We assume that $\vect{\Sigma}_K$ is a diagonal matrix, with elements
\begin{equation}
\vect{\Sigma}_K = \mrm{diag}( l^2_{\Om}, l^2_{S_8}, l^2_{h}, l^2_{\fr0}, l^2_{M_{200}} ),
\label{eq:kernel_covariance}
\end{equation}
where (for example) $l_{S_8}$ is a length-scale in terms of the parameter $S_8$. The intuition one should have is that our prior will substantially down-weight functions where $g_i$ fluctuates as $S_8$ is varied, if the length-scale of the fluctuations is shorter than $l_{S_8}$. The value of $\sigma_K$ determines the prior expectation on the amplitude of variations in $g_i$ as one varies $\vect{\theta}$.

We build a separate Gaussian process to emulate each of the 24 GIK parameters. For a single GIK parameter, $g_i$, the inputs to our Gaussian process are:
\begin{itemize}
\item The estimates of $g_i$ evaluated for each of our 7 cluster mass bins, at each of the 50 FORGE cosmologies ($\hat{g}_i(\vect{\theta}_j)$), for which we use the mean of the marginalised posterior on this particular GIK parameter.
\item The Gaussian uncertainty on each $\hat{g}_i(\vect{\theta}_j)$, which we calculate as the square root of the variance of the relevant marginalised posterior.
\item A kernel function (eq.~\ref{eq:kernel_function}), with its associated hyperparameters.
\item A mean function, which we assume to be a constant (i.e. independent of $\vect{\theta}$), and which we set equal to $\left< g_i(\vect{\theta}_j) \right>_{j}$, the mean value of $g_i$ over the 350 ``cosmologies'' where the GIK parameters have been fit.
\end{itemize}

\subsection{Selecting kernel hyperparameters}

For each of our GIK parameters, we have six kernel hyperparameters ($\sigma_K$ in eq.~\ref{eq:kernel_function}, and the five length-scales in eq.~\ref{eq:kernel_covariance}). In order to set these hyperparameters in an objective manner, we find (separately for each $i$) the set of hyperparameters that maximise the marginal likelihood of the input data \citep[see equation~5.8 of][]{2006gpml.book.....R}. Considering the combination of our kernel function (with a specific set of hyperparameters) and mean function as defining a prior on the function $g_i(\vect{\theta})$, the marginal likelihood is the probability density associated with having drawn our input data from this prior, and can be calculated by \george.

We note that some work in the literature alternatively determines the optimal hyperparameters using \emph{leave one out} tests \citep[called ``cross-validation'' in][]{2006gpml.book.....R}. This involves building one Gaussian process per piece of training data, where each Gaussian process is trained on all but one, \emph{left out}, piece of training data. Each Gaussian process is then used to predict the value of the relevant left out training data. This can be done for a variety of different kernel hyperparameters, with the predicted and true values of the left out data compared, and the hyperparameters that lead to the most accurate predictions for the left out data chosen. We implemented such a method, and found best-fit hyperparameters that were similar to those from maximising the marginal likelihood. As this method is slow (it requires building a separate Gaussian process for each piece of training data) we used the marginal likelihood approach for this work.

\section{Fitting the emulator model to mock $\xicg$ data}
\label{sect:fitting_to_mock_data}

Combining an emulator for the GIK parameters as a function of the cosmological parameters, with knowledge of the real-space cluster-galaxy correlation function, we are able to predict $\xicg$ as a function of cosmology. By comparing these model predictions with some observed $\xicg$ data vector, we can then constrain the cosmological parameters. In order to make this comparison, we need to define a likelihood of having obtained a particular observed $\xicg$, given a model $\xicg$. We assume this likelihood is Gaussian, with
\begin{equation}
\mathcal{L}(\vect{\theta}) \propto \exp \left( - \frac{1}{2} (\xicgd - \xicg(\vect{\theta}))\transpose \vect{C}^{-1} (\xicgd - \xicg(\vect{\theta})) \right),
\label{eq:likelihood}
\end{equation}
where $\xicgd$ is the observed data vector, $\xicg(\vect{\theta})$ is the model-predicted data vector for the cosmological parameter vector $\vect{\theta}$, and $\vect{C}$ is the data covariance matrix.

\subsection{Covariance matrix for $\xicg$}
\label{sect:cov}

Even if adopting the correct cosmological parameters, with a perfect cosmological model, we would not expect the model data vector to match the observed one. Instead, we would expect the observed data vector to look like a random draw from a multivariate Gaussian distribution that has a mean value equal to the model data vector and some covariance matrix.\footnote{This is what is meant by assuming that our ``likelihood is Gaussian''.} This covariance matrix describes the amount of noise on each element of the observed data vector, as well as correlations between the noise on different elements of the data vector. If we could generate an infinite number of independent realisations of a Gaussian distributed data vector, $\vect{d}$, then the covariance matrix for $\vect{d}$ would be $C_{ij} \equiv \left< \, (d_i - \left< d_i \right> )(d_j - \left< d_j \right> ) \, \right> $, where $\left< \vect{d} \right>$ is the mean of the data vectors, and $d_i$ is the $i$th element of $\vect{d}$.

To estimate a covariance matrix suitable for our purposes, we use the Indra simulations that were described in Section~\ref{sect:Indra}. The data vector we use when fitting for the cosmological parameters is $\xicg(r_p, r_\pi)$ on a $15 \times 20$ grid of $(r_p, r_\pi)$, with $(1 \, \hmpc)^2$ pixels, meaning that we have a data vector with $n_\mathrm{d} = 300$ elements.
For the purposes of discussing the covariance matrix, it is simpler to consider our data vector to be one dimensional. In order to facilitate this, we flatten our 2D data vector, such that elements 1 to 20 run from small to large $r_\pi$ with $r_p = 0.5 \, \hmpc$, elements 21 to 40 are the same at $r_p = 1.5 \, \hmpc$, etc. We label this flattened version of the data vector $\vect{\xi}$, with $\xi_i$ the $i$th element, with $i$ running from 1 to 300.

In total there are $n_\mathrm{s} = 384$ Indra simulations, each a $1 \, \hgpc$ on-a-side box, all run with the same cosmological parameters. We can use each Indra simulation to calculate a redshift-space cluster-galaxy correlation function data vector, $\vect{\xi}^k$, with $k$ indicating the index of the Indra simulation. These can then be used to produce an unbiased estimate for the data covariance matrix
\begin{equation}
\widehat{C}_{ij} = \frac{1}{n_\mathrm{s} - 1} \sum_{k=1}^{n_\mathrm{s}} \left(\xi_i^k - \left< \xi_i \right> \right) \left(\xi_j^k - \left< \xi_j \right> \right),
\label{eq:data_cov}
\end{equation}
where the $n_\mathrm{s} - 1$ factor in the denominator (as opposed to simply $n_\mathrm{s}$) accounts for the fact that we have estimated the mean $\vect{\xi}$ from the data itself.

While $\widehat{\vect{C}}$ is an unbiased estimate of $\vect{C}$, $\widehat{\vect{C}}^{-1}$ is not an unbiased estimate of the \emph{precision matrix} $\vect{\Psi} \equiv \vect{C}^{-1}$. However, we can achieve an unbiased estimate of $\vect{\Psi}$ as \citep{2007A&A...464..399H}
\begin{equation}
\widehat{\vect{\Psi}} = \frac{n_\mrm{s} - n_\mrm{d} - 2}{n_\mathrm{s} - 1} \widehat{\vect{C}}^{-1}.
\label{eq:hartlap}
\end{equation}
Note that at a minimum, one requires $n_\mathrm{s} > n_\mathrm{d} + 2$ in order for $\widehat{\vect{C}}$ to be invertible. However, if $n_\mathrm{s}$ is only slightly larger than $n_\mathrm{d} + 2$, while equation~\ref{eq:hartlap} gives an unbiased estimate of the precision matrix, the uncertainty on this estimate will be large \citep{2013MNRAS.432.1928T}. One generally requires that $n_\mathrm{s}$ is considerably larger than $n_\mrm{d}$ in order for $\vect{\Psi}$ to be precisely determined. In our case we have a length-300 data vector, with only 384 simulations, which would lead to significant uncertainty on the covariance matrix.

Without access to a larger number of simulations, or analytic estimates for the covariance matrix, we can instead turn to resampling techniques. These techniques include the \emph{sub-sample}, \emph{bootstrap} and \emph{jackknife} methods \citep{2009MNRAS.396...19N}. They are usually applied when fitting to an observed data vector, and involve using the data itself (split up into sub-regions) to estimate the data covariance. Similar techniques can also be used to decrease the number of independent simulations required to compute a covariance matrix to some prescribed level of accuracy. For example, \citet{2016arXiv160600233E} performed Jackknife resampling over a set of independent simulations, finding that this reduced the requirement on the number of simulations (in their case) by a factor of 7.

We follow \citet{2016arXiv160600233E} and calculate a covariance matrix from each of our individual simulations, using the delete-$d$ Jackknife scheme \citep{10.1214/aos/1176347263}.\footnote{We note that a similar strategy was used in \citet{2022MNRAS.517..374H}.} We split each simulation into $N_s=8$ subregions, by splitting the box in half along each of the three Cartesian axes. Each jackknife configuration, $c$, then leaves out $N_d=2$ of the $N_s$ regions, and we evaluate the corresponding correlation function, $\vect{\xi}^c$, by adopting the simulation $z$-axis as the line-of-sight direction, and using the cluster and galaxy definitions given in Section~\ref{sect:Indra}. Note that whether a particular cluster-galaxy pair was included in a Jackknife region was based upon the position of the cluster, such that clusters close to the boundary of a ``deleted'' subregion, can appear in pairs with galaxies from the deleted region. The covariance matrix estimate from each simulation is 
\begin{equation}
    \widehat{C}_{ij}^{1 \mrm{sim}} =\frac{N_s-N_d}{N_d \, N_{\textsc{jk}}} \sum\limits_{c=1}^{N_{\textsc{jk}}}  \left(\xi_i^c - \left< \xi_i \right> \right) \left(\xi_j^c - \left< \xi_j \right> \right), 
 \label{eq:cov_matrix_JK}
\end{equation}
where $N_{\textsc{jk}}= { N_s \choose N_d}$ is the number of possible jackknife configurations. Having evaluated a separate $\widehat{\vect{C}}^{1 \mrm{sim}}$ from each Indra simulation, our estimate for the data covariance matrix is the average of all of these, and our estimate for the inverse data covariance matrix is just the inverse of this average \citep[without any factor like that in equation~\ref{eq:hartlap}, following][]{2016arXiv160600233E}.


The covariance matrix evaluated in this way is appropriate for the $(1 \, \hgpc)^3$ volume of an Indra simulation. Where necessary, we assume that the covariance matrix scales inversely with volume \citep[e.g.][]{2017MNRAS.472.4935H}, in order to calculate a covariance matrix appropriate for other volumes.

\begin{figure*}
        \centering
        \includegraphics[width=0.8\textwidth]{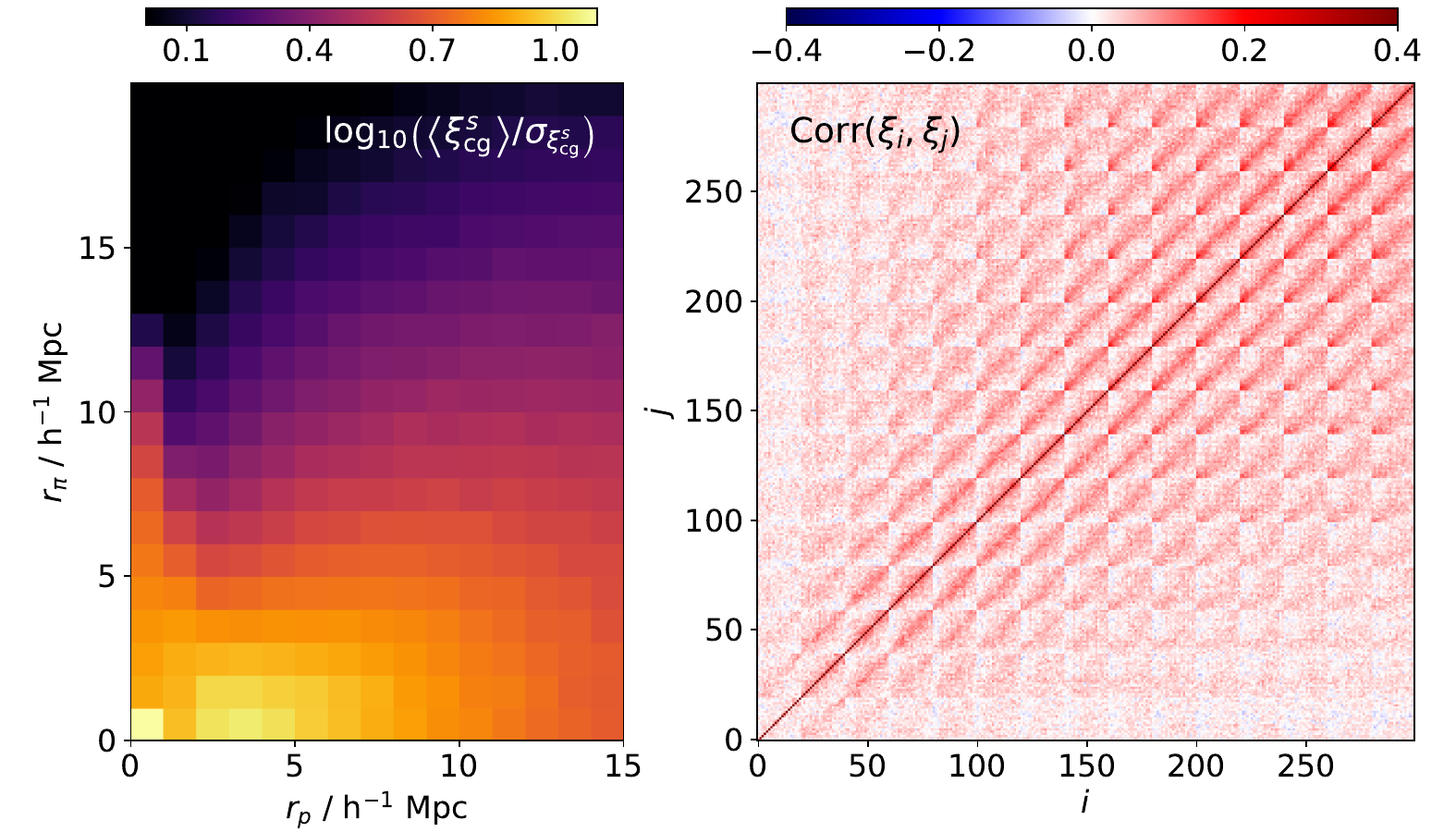}     
	\caption{Left: a map of the signal-to-noise ratio (SNR) in the Indra $\xicg$ maps. This is calculated by dividing (pixel-by-pixel) the mean value of $\xicg$ from the various Indra sub-volumes by the standard deviation of $\xicg$ over the sub-volumes. Note that the values of $\sigma_{\xicg}^2$ are equal to the diagonals of the covariance matrix, $C_\mathrm{Indra}$. Right: the correlation matrix, corresponding to the covariance matrix, $C_\mathrm{Indra}$. The values along the diagonal are (by definition) all equal to 1, but we use a smaller range for the colour map to highlight the correlations between many of the pixel values.}
	\label{fig:cov_matrix}
\end{figure*}

In Fig.~\ref{fig:cov_matrix} we plot the correlation matrix, which is closely related to the covariance matrix just described. The correlation matrix measures the Pearson correlation coefficient between all pairs of elements, and is defined as $\mathrm{Corr}(\xi_i, \xi_j) \equiv \widehat{C}_{ij} / \sqrt{\widehat{C}_{ii} \widehat{C}_{jj}}$. In the right panel of Fig.~\ref{fig:cov_matrix} one can see that neighbouring pixels of $\xicg(r_p, r_\pi)$ are quite correlated,\footnote{Given the order in which we flatten $\xicg(r_p, r_\pi)$ to get $\vect{\xi}$, neighbouring indices correspond to neighbouring pixels in the $r_\pi$ direction, while indices separated by 20 correspond to neighbouring pixels in the $r_p$ direction.} suggesting that this particular data vector may not be the most efficient compression of the set of small-scale $(r_p, r_\pi)$ separations for cluster-galaxy pairs. One could instead consider \emph{multipoles} of the correlation function \citep[e.g.][]{2019ApJ...874...95Z}, \emph{clustering wedges} \citep[e.g.][]{2012MNRAS.419.3223K}, or different numbers of pixels and/or different spacing of bin edges \citep[such as the logarithmically-spaced $r_p$ bin edges in][]{2022MNRAS.515..871Y}. An advantage of these alternatives is that they generally lead to a smaller number of data vector elements, decreasing the requirements on the number of simulations required to evaluate the covariance matrix. However, we leave investigating these alternatives to future work.

In the left panel of Fig.~\ref{fig:cov_matrix} we plot a map of the signal-to-noise ratio (SNR) from the ensemble of Indra simulations. The signal in this case is the mean value of $\xicg$ over the different Indra sub-volumes, while the noise is the square root of the diagonal elements of $\widehat{\vect{C}}$, which we label $\sigma_{\xicg}$. Both the signal and noise increase as one approaches the origin (i.e. $r_p = r_\pi = 0$), but the SNR peak is at low-$r_\pi$, with $r_p \sim 4 \, \hmpc$.

\subsection{Including model uncertainty in the covariance matrix}
\label{sect:model_uncertainty}

The covariance matrix calculated from the Indra simulations, which we here label $C_\mathrm{Indra}$, describes how large the variations typically are between $\xicg$ calculated from any one particular Indra simulation, and the average of all the Indra simulations (which, given the large number of Indra simulations, we can think of as being a good approximation to the ``noise free'' prediction for $\xicg$ at the Indra cosmology). 

This means that if our model for $\xicg(\vect{\theta})$ could produce the ``noise free'' prediction for $\xicg$ as a function of cosmological parameters, we could set $C$ in equation~\eqref{eq:likelihood} equal to $C_\mathrm{data} = C_\mathrm{Indra} / (V_\mathrm{data} / V_\mathrm{Indra})$, with $V_\mathrm{data}$ the effective volume of our survey, and $V_\mathrm{Indra} = (1 \, \hgpc)^3$. However, our emulator for the GIK parameters is built from simulations with a finite volume, which leads to uncertainty on the GIK parameters (which can be visualised, for example, by the shaded regions around the GIK function lines in Fig.~\ref{fig:ZW_Fig5}). 




Recall that we have a separate Gaussian Process emulator for predicting each element, $g_i$, of the GIK parameter vector, $\vect{g}$, as a function of $\vect{\theta}$, and that these predictions take the form of a Gaussian probability distribution, specified by a mean prediction $\left< g_i (\vect{\theta}) \right>$, and a standard deviation $\sigma_{g_i}(\vect{\theta})$. This can also be expressed as having a predicted mean GIK parameter vector, $\left< \vect{g} (\vect{\theta})\right>$, and a covariance matrix for $\vect{g} (\vect{\theta})$, which we label $\vect{\Sigma}_g$. For now $\vect{\Sigma}_g$ is diagonal (with the diagonal elements equal to the $\sigma^2_{g_i}(\vect{\theta})$), because we have built a separate emulator for each element of $\vect{g}$, and therefore cannot account for any covariance between the different elements of $\vect{g}$.


\subsubsection{``Diagonalised'' GIK parameters} 
\label{sect:diagonalise}

While there has been work on how to formulate \emph{multi-output} Gaussian Processes \citep[e.g][]{WANG2015159, 2017arXiv170901298P}, we are not aware of any available Gaussian Process software that can take vector-valued training data such as the values of $\vect{g}(\vect{\theta}_j)$ for a set of $\vect{\theta}_j$, along with correlated uncertainties on each element of $\vect{g}(\vect{\theta}_j)$, and then make predictions for $\vect{g} (\vect{\theta})$ along with $\vect{\Sigma}_g(\vect{\theta})$. By building a separate emulator for each element of $\vect{g}$, we implicitly ignore the off-diagonal elements of the uncertainty on the measurements of the $\vect{g}(\vect{\theta}_j)$, and force $\vect{\Sigma}_g$ to be diagonal.

Given that the posterior distributions on the $\vect{g}(\vect{\theta}_j)$ do have quite considerable degeneracies between the different elements of $\vect{g}$, it would be better not to ignore them. We can circumvent the requirement for implementing a multi-output Gaussian process, by finding a linear transformation, $\vect{g'} = \vect{L} \vect{g}$, such that the posterior distribution for $\vect{g'}$ is approximately a separable function of each element of $\vect{g'}$, which is to say that the posterior for each $\vect{g'}$ in our set of training data has little covariance between the different elements.

We define the covariance matrix for a GIK parameter posterior distribution as 
\begin{equation}
\vect{M} = \left< \, (\vect{g} - \left< \vect{g} \right> )(\vect{g} - \left< \vect{g} \right> )\transpose \, \right>,
\end{equation}
where the expectation values can be estimated from an MCMC chain of GIK parameter vectors. Then, defining $\vect{V}$ as a matrix whose columns are the unit-length eigenvectors of $\vect{M}$, we have that $\vect{g'} = \vect{V}\transpose \vect{g}$.

The idea then, is that because each piece of training data (in terms of $\vect{g'}$) has only minimal covariance between the different elements, that building a separate Gaussian process emulator for each element of $\vect{g'}$ (rather than $\vect{g}$), and propagating the uncertainties on the training data $\vect{g'}$s through the Gaussian process, allows us to more faithfully include the covariance in our training data between the different elements of $\vect{g}$. The Gaussian processes then return an estimate of $\vect{g'}$ along with a diagonal covariance matrix, $\vect{\Sigma}_{g'}$, and from this we can evaluate $\vect{g} = \vect{V} \vect{g'}$ and $\vect{\Sigma}_{g} = \vect{V} \vect{\Sigma}_{g'} \vect{V}\transpose$ (which can now have non-zero off-diagonal elements, representing the covariance between the different elements of $\vect{g}$).

\subsubsection{Mapping GIK parameter uncertainties into uncertainties on $\xicg$}

In order to evaluate the likelihood for a particular $\vect{\theta}$ (equation~\ref{eq:likelihood}), we need to find the GIK model-predicted $\xicg(\vect{\theta})$, which is done following the procedure described in Section~\ref{sect:calc_xicg_with_GIK}. What we would then like to know is how uncertainty on $\vect{g}(\vect{\theta})$ (described by the covariance $\vect{\Sigma}_{g}$), maps in to uncertainty on $\xicg(\vect{\theta})$, such that this uncertainty on the model prediction for $\xicg(\vect{\theta})$ can be accounted for in our covariance matrix. To do this, we consider small perturbations to the GIK parameter vector about some reference value, $\vect{g}_0$. We assume that the correlation function in such a case can be well approximated by
\begin{equation}
\vect{\xi} \approx \vect{A} (\vect{g} - \vect{g}_0) + \vect{\xi}_0,
\end{equation}
where $\vect{\xi}_0$ is the data vector evaluated for the GIK parameter vector $\vect{g}_0$. The matrix $\vect{A}$ is the gradient of $\vect{\xi}$ with respect to $\vect{g}$, $A_{ij} = \partial \xi_i / \partial g_j$. This can be calculated using a finite difference approach, by evaluating $\vect{\xi}(\vect{g})$ over a grid of $\vect{g}$-points around $\vect{g}_0$. The covariance matrix for $\vect{\xi}$ due to uncertainties on $\vect{g}$ is then $\vect{C}_g = \vect{A} \vect{\Sigma}_{g} \vect{A}\transpose$, where we evaluate $\vect{A}$ at a fiducial cosmology (and treat it as independent of cosmology), while $\vect{\Sigma}_{g}(\vect{\theta})$ is returned by our Gaussian Processes, alongside $\vect{g}(\vect{\theta})$.

We note that not doing our diagonalisation procedure would be a conservative choice. This is because the GIK parameter posteriors can have quite tight parameter degeneracies. If we consider two highly degenerate parameters, then their individual marginal distributions can be quite broad, while some combination of them is well constrained. If we ignore the fact that there is a well-constrained combination, and instead set their joint distribution equal to the product of the two marginals, then this creates more freedom for the parameters to take different values, and therefore more uncertainty on the GIK functions, and a resulting increased model uncertainty on $\xicg$. 

To demonstrate this quantitatively, we calculated a $\chi^2$ between a fiducial $\xicg$ (from $\vect{g}(\vect{\theta})$ at the mean FORGE cosmology), and the $\xicg$ evaluated for the best-fit GIK parameters for each of the 50 FORGE simulations. Doing this with and without ``diagonalisation'', we found that the mean $\chi^2$ without diagonalisation is $\sim$70\% of that with diagonalisation.



\subsection{Fitting to mock data generated by the emulator}

As a first demonstration of using our emulator to fit to a $\xicg$ data vector, we fit to a data vector generated by the emulator itself. Such an exercise only partially tests our emulation procedure, but has the advantage that the data vector being fit to can be generated ``noise free'', or with some other level of noise that we specify. This differs from a data vector constructed from an $N$-body simulation, which will necessarily have a level of noise set by the finite volume of the simulation.

\begin{figure}
        \centering
        \includegraphics[width=\columnwidth]{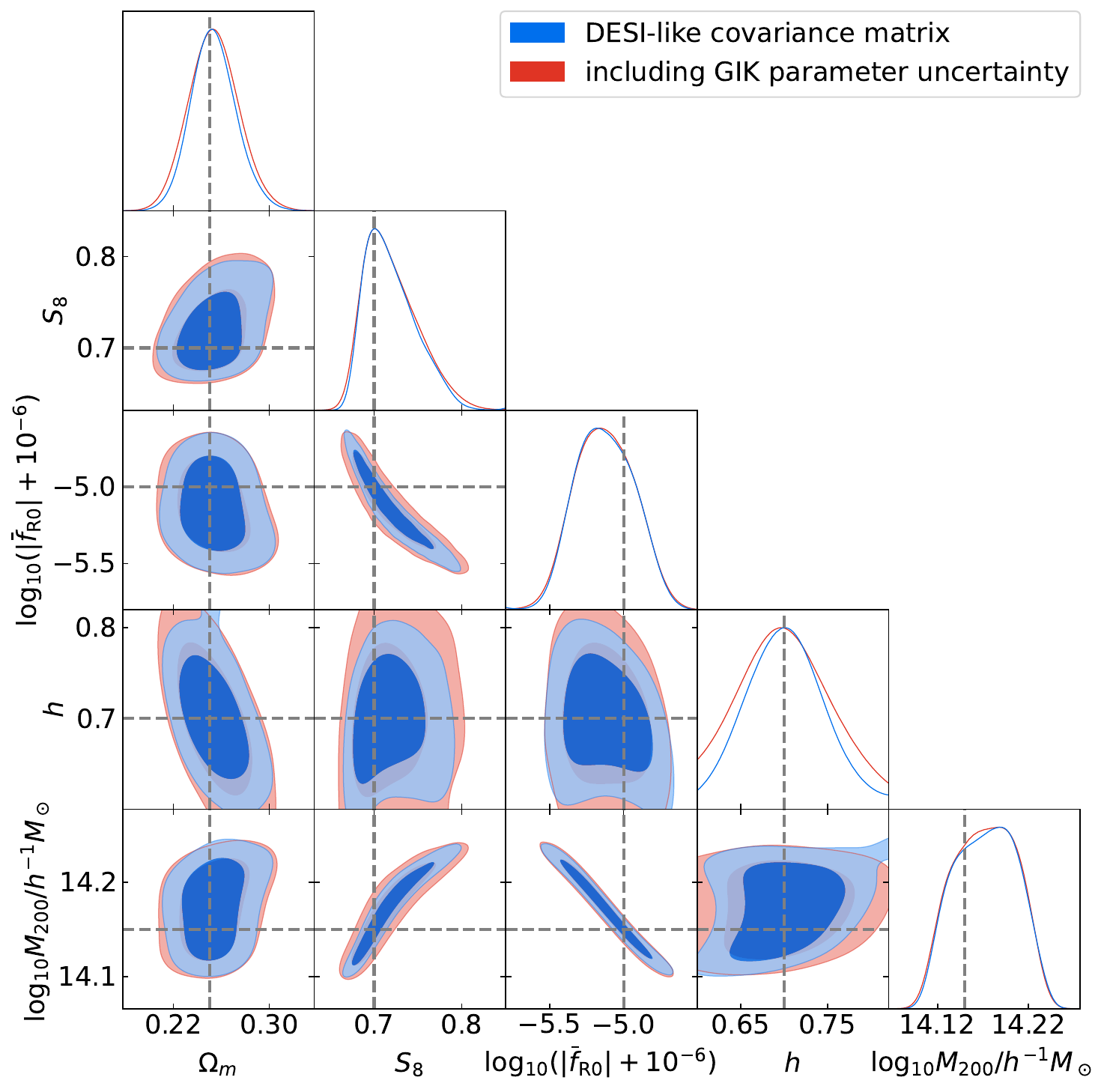}    
	\caption{The posterior distribution for the cosmological parameters when fitting to a data vector generated by our emulator. The input cosmology is marked by the dashed lines, and the blue contours contain 68 and 95\% of the posterior distribution when using a covariance matrix (within our likelihood) that corresponds to the Indra data covariance re-scaled for an effective volume of $20 \, (\hgpc)^3$. The red contours show the results when adding terms to the covariance matrix to account for the fact that we only know the GIK parameters as a function of cosmology to some finite level of precision, which is described in Section~\ref{sect:model_uncertainty}.}
	\label{fig:DESI_forecast}
\end{figure}

In Fig.~\ref{fig:DESI_forecast} we show the cosmological constraints from such an exercise. We build an emulator using the 50 FORGE simulations, and then use the emulator to predict a model data vector at a particular point in our cosmological parameter space (marked by the vertical and horizontal dashed lines in Fig.~\ref{fig:DESI_forecast}). We then use MCMC to sample from the posterior distribution for $\vect{\theta}$, with flat priors on each element of $\vect{\theta}$ (see eq.~\ref{eq:cosmo_theta}) covering the range of $\vect{\theta}$ over which FORGE simulations were run, and using a data covariance matrix that is equal to that calculated for $(1 \, \hgpc)^3$ from the Indra simulations, but scaled down by a factor of 20, such that it is approximately appropriate for the effective volume of a survey such as DESI \citep{2016arXiv161100036D}, which we take to have an effective volume of $20 \, (\hgpc)^3$ following \citet{2022MNRAS.515.1854G}.

Fig.~\ref{fig:DESI_forecast} reveals some interesting degeneracies between the different cosmological parameters. Perhaps the most striking is the degeneracy between the cluster mass associated with our cluster sample and $\fr0$. In particular, a large cluster mass with a low value for $\log |\fr0|$, produces a similar $\xicg$ to a smaller cluster mass with a higher value for $\log |\fr0|$. Given that we expect the gravitational attraction of more massive clusters to be stronger, and that a larger $\log |\fr0|$ means an increased enhancement in the gravitational forces, this degeneracy makes intuitive sense, in that an enhancement to gravity can make clusters appear more massive than they actually are. It also suggests that combining the modelling of $\xicg$ with some independent measurement of cluster masses will be a fruitful avenue for constraining the nature of gravity. For example, weak lensing can measure the mass of clusters in a manner that is unaffected by $\fr0$ (because the modifications to gravity do not affect the propagation of photons in $f(R)$ gravity). Such a measurement would break the $M_{200} - \fr0$ degeneracy seen in Fig.~\ref{fig:DESI_forecast}, leading to an improved constraint on $\fr0$.


Another feature of Fig.~\ref{fig:DESI_forecast} is that we can see the impact that including the GIK parameter uncertainty into the covariance matrix (following Section~\ref{sect:model_uncertainty}) has on the cosmological constraints. In particular, while not dominant over the data covariance, the constraints are broadened somewhat when accounting for the GIK parameter uncertainty. This means that for a DESI-like survey one would ideally want larger volume simulations than the FORGE simulations we used here, in order for this model uncertainty to be negligible for the derived constraints. We stress that the ability for $(500 \, \hmpc)^3$ volume simulations to be used in the modelling of a $20 \, (\hgpc)^3$ data vector, without the model uncertainty being dominant over the data covariance, is possible only because of employing the GIK model (as opposed to directly emulating the data vector extracted from the set of simulations) and the resulting advantages described in Section~\ref{sect:pros_cons_vs_direct_xicg}.


\subsection{Fitting to mock data extracted from simulations}

As described above, while fitting to emulator-generated data provides some tests of our analysis pipeline, and also acts as a forecast for the constraining power expected with a particular experiment (which enters through the covariance matrix), it does not provide an exhaustive test of our modelling procedure. For example, we have seen in Fig.~\ref{fig:Pvrvt_example} that the functional form that we use for $P(v_r,v_t | r)$ does not perfectly describe the galaxy-cluster pairwise velocity distribution. To assess the extent to which this mismatch could lead to biased results, we also want to test fitting to data vectors measured directly from simulations (as in, to ``observe'' our simulations much like we observe the real Universe).

\subsubsection{Stacked Indra $\xicg$}
\label{sect:fit_to_stacked_Indra}

One problem with testing on data vectors measured from simulations, is that simulations with sufficient resolution that we can populate them with galaxies using a subhalo abundance matching-like scheme are typically not of comparable volume to future surveys. This means that while we can test that the model gives reasonable results with a modest volume of data, we cannot test that it returns results that are unbiased at the level of precision that will be afforded by future datasets.

To get around this, we return to the Indra suite of simulations, which comprises many simulations, each of which has a $(1 \, \hgpc)^3$ volume. These simulations were all run with the same cosmology, and so we can combine the $\xicg(r_p, r_\pi)$ from many of them to get a simulated datavector, at the Indra cosmology, from a large simulated volume.\footnote{We note that this is not quite equivalent to having run a single large volume simulation, primarily due to the absence of density fluctuations with wavelengths larger than the boxsize in each of the Indra simulations.} Specifically, we calculate $\xicg(r_p, r_\pi)$ from each Indra simulation (using our fiducial cluster and galaxy definitions), and then take the mean of this over the different Indra simulations to get our \emph{stacked Indra} data vector (this is the same as $\left< \xi^{s,k}_\mrm{sim} \right>$ from Section~\ref{sect:pros_cons_vs_direct_xicg}).

\begin{figure}
        \centering
        \includegraphics[width=\columnwidth]{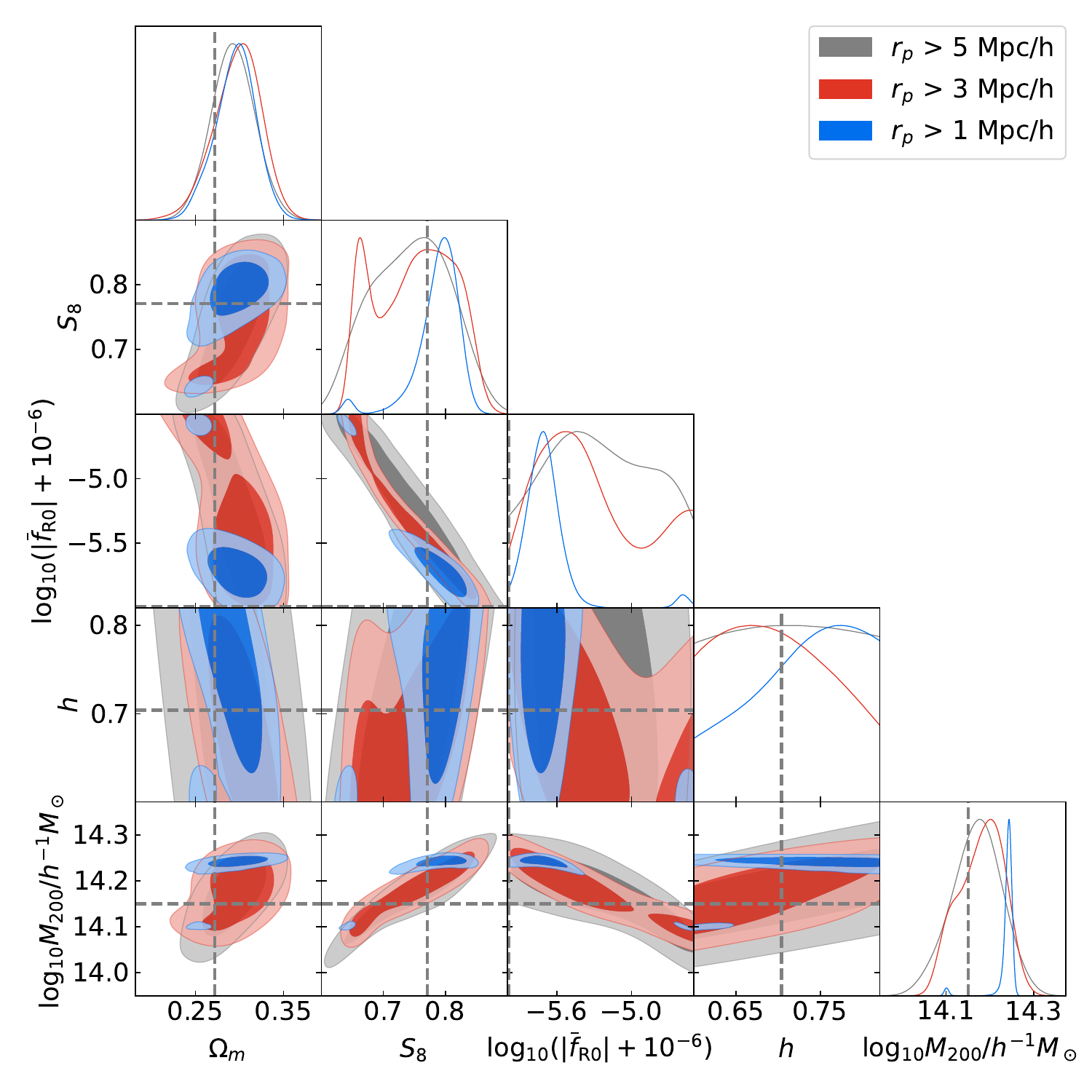}  
	\caption{The posterior distribution for the cosmological parameters when fitting to a data vector calculated from a stack of the Indra simulations. The different colours correspond to fitting the $\xicg$ data vector down to different minimum $r_p$. The Indra cosmology is marked by the dashed lines. Because the Indra simulations were run with GR, they have $\bar{f}_\mrm{R0}=0$, corresponding to $\log_{10} |\bar{f}_\mrm{R0} + 10^{-6}| = -6$, which is on the edge of our cosmological parameter prior.}
	\label{fig:stacked_Indra_fit}
\end{figure}

We fit to this stacked Indra data vector using MCMC, with the model  $\xicg(\vect{\theta})$ calculated from our FORGE-trained emulator for the GIK parameters, along with the true real-space clustering (for which we use the mean over the different Indra simulations of $\xi_\mathrm{cg}^r(r)$). The resulting posterior distribution is shown in Fig.~\ref{fig:stacked_Indra_fit}, where the different colours correspond to different minimum $r_p$ values down to which we fit the data. 

A key takeaway from Fig.~\ref{fig:stacked_Indra_fit} is that if we make our fiducial choice of fitting to $\xicg$ over the range $1 \leq r_p / \hmpc \leq 15$, $0 \leq r_\pi / \hmpc \leq 20$, then we confidently infer a cosmology that is inconsistent with the true cosmology used to run the Indra simulations. This inconsistency goes down -- both due to a broadening of the posterior, and due to a shifting of the mean of the posterior -- if we increase the minimum value of $r_p$ down to which we fit the data. The fact that using the GIK model to predict $\xicg$ on scales $\lesssim 3 \, \hmpc$ can lead to biases in the inferred cosmology should not come as a surprise given the evidence for model misspecification that we saw on these scales in Fig.~\ref{fig:GIK_vs_sim}. 

It is interesting to consider why the constraints are noticeably tighter when one includes scales down to $1 \, \hmpc$. This suggests that there is considerable cosmological information contained in the velocities of galaxies $1 - 3 \, \hmpc$ away from clusters. We hypothesize that this is because the radius at which $f(R)$'s Chameleon screening turns on (and so inside of which, the usual GR forces are obtained) is of order a Mpc. For example, at $z=0$, and in a halo slightly less massive than the clusters we consider here, the screening radius is between $1$ and $2 \mpc$ when $\log_{10} |\bar{f}_\mrm{R0}| = -6$ \citep[see Fig.~7 of][]{2015MNRAS.448..307C}. If some enhancement to gravity happens everywhere, then it is hard to distinguish between a more massive cluster and enhanced gravity. However, with data that spans regions that -- with $f(R)$ -- would be both screened and unscreened, it is no longer the case that enhancing gravity (in only the unscreened regions) is equivalent to increasing the cluster mass. 

We should stress that the GIK parameter emulator used with these different minimum $r_p$ values was subtly different. We had originally fit the GIK parameters for each FORGE simulation over the radial range $1 < r / \hmpc < 20$, and used these GIK parameter posteriors when training our GIK parameter emulator. However, we found that when only fitting to the $\xicg$ data vector at $r_p > 3 \, \hmpc$, we got a less biased result if our GIK parameter emulator was trained using GIK parameters fit to the radial range $3 < r / \hmpc < 20$ rather than $1 < r / \hmpc < 20$. This makes sense, because if the GIK parameters returned by fitting to $r > 1 \, \hmpc$ and $r > 3 \, \hmpc$ are different, the $r > 3 \, \hmpc$ case must be a better description of the cluster--galaxy pairwise velocity distribution at large radii. Forcing the same set of GIK parameters to simultaneously explain the velocity distribution within $1 < r / \hmpc < 3$ (where we know the model is not an accurate description of the velocity distribution) then degrades how well the GIK model describes the velocity distribution at large radii. Given that $\xicg$ at some $r_p$ is affected only by the clustering and velocity distribution at 3D radii $r \geq r_p$, if we only need to model the data vector at $r_p > r_\mathrm{min}$, then it makes sense to only fit the GIK parameters to radii $r > r_\mathrm{min}$.

An avenue for future work would be to improve the GIK model such that it provides a better description of $P(v_r,v_t | r)$ at small $r$. This could involve both improving upon the 7-parameter model for the velocity distribution at a single radius (equation~\ref{eq:model_vel_distribution}), which we have seen in Fig.~\ref{fig:Pvrvt_example} is not a perfect description of the velocity distribution found in $N$-body simulations, as well as finding improved functional forms for how these 7 GIK functions vary with radius (e.g. the form of equation~\ref{eq:my_fr}).

For now, we note that while our GIK model-based emulator allows for the extraction of cosmological information from RSD on scales below that at which linear theory predictions break down, it does not appear that we can use it to model cluster--galaxy clustering, all the way down to scales below the virial radii of the clusters. This is somewhat disappointing given that one of the key features of the GIK model is that the velocity distribution has two components, one of which describes galaxies that are virialised within galaxy cluster potentials (and which is only relevant on scales around and below cluster virial radii).

Another potential explanation for the biased cosmological parameter inference seen in Fig.~\ref{fig:stacked_Indra_fit} is differences between the FORGE and Indra simulations. If the Indra and FORGE simulations predicted different GIK parameters for the same cosmology, then using a FORGE-trained emulator to fit to Indra mock data would lead to such a bias.\footnote{This would also suggest that the GIK parameters are not robustly predicted by simulations, which would be bad for the prospects of applying this model to observational data.} While this effect is almost certainly present at some level, the reasonable convergence of the GIK functions with respect to numerical resolution seen in Fig.~\ref{fig:TNG_ZW_Fig5} suggests that the differences between Indra and FORGE (most notably that Indra has eight times worse mass resolution) should not lead to dramatic differences.

To be more quantitative about this, we inspected the GIK parameters predicted by our FORGE-trained emulator, at the Indra cosmology (there is no FORGE simulation run at the Indra cosmology). We define the following:
\begin{itemize}
\item $\vect{g}_\mathrm{FORGE}(\vect{\theta}_\mathrm{Indra})$, the GIK parameters predicted by the FORGE-trained emulator, at the cosmology of the Indra simulations.
\item $\vect{g}^k_\mathrm{Indra}$, the GIK parameters measured for the $k$th Indra simulation.
\item $\left< \vect{g}^k_\mathrm{Indra} \right>$, the mean of the GIK parameters from the different Indra simulations.
\item $\vect{\theta}_{\xicg}$, the maximum-likelihood cosmological parameters when fitting to the stacked-Indra $\xicg$ data vector (the ``$r_p > 1 \, \hmpc$'' fit in Fig.~\ref{fig:stacked_Indra_fit}).
\item $\vect{g}_\mathrm{FORGE}(\vect{\theta}_{\xicg})$, the GIK parameters predicted by the FORGE-trained emulator, at $\vect{\theta}_{\xicg}$. These are the GIK parameters being used in the best-fitting model to the stacked-Indra $\xicg$.
\item $\vect{\Sigma}_\mathrm{g}^\mathrm{Indra}$, the covariance matrix for $\vect{g}$ calculated from the set of $\vect{g}^k_\mathrm{Indra}$.
\end{itemize}
Then, defining $\chi^2(\vect{g}_a,\vect{g}_b)$ as $(\vect{g}_a - \vect{g}_b)\transpose (\vect{\Sigma}_\mathrm{g}^\mathrm{Indra})^{-1} (\vect{g}_a - \vect{g}_b)$, we find that  $\chi^2(\vect{g}_\mathrm{FORGE}(\vect{\theta}_\mathrm{Indra}), \left< \vect{g}^k_\mathrm{Indra} \right>) = 177$, while $\chi^2(\vect{g}_\mathrm{FORGE}(\vect{\theta}_{\xicg}), \left< \vect{g}^k_\mathrm{Indra} \right>) = 571$. So the difference between $\vect{\theta}_{\xicg}$ and $\vect{\theta}_\mathrm{Indra}$ is not because $\vect{g}_\mathrm{FORGE}(\vect{\theta}_{\xicg})$ does a better job than $\vect{g}_\mathrm{FORGE}(\vect{\theta}_\mathrm{Indra})$ of producing GIK functions that look like those found in Indra. This points towards the dominant effect being model misspecification, that results in the optimal set of GIK parameters to describe $\xicg$ not being the same as those one gets by fitting the GIK parameters directly to the position/velocity data from simulations.


Nevertheless, $\chi^2(\vect{g}_\mathrm{FORGE}(\vect{\theta}_\mathrm{Indra}), \, \left< \vect{g}^k_\mathrm{Indra} \right>) = 177$ (compared with an expectation of $\sim 24$ if the FORGE prediction at the Indra cosmology looked like the GIK parameters from a random Indra volume) suggests there are systematic differences between Indra and FORGE, and/or errors induced by the emulation process. In order to get a sense of how robust the GIK parameters are to the details of the simulations, in the future it would be good to compare the GIK parameters fit to different simulations run with the same cosmology, and also to build emulators with different suites of simulations and compare their results.







\section{Discussion}
\label{sect:discussion}

The procedure that we have presented, combining an emulator for the cluster--galaxy pairwise velocity distribution with some external knowledge (or model) for the real space clustering, is an interesting alternative to more direct emulation of redshift-space clustering data vectors from simulations. As described in Section~\ref{sect:pros_cons_vs_direct_xicg} it has distinct advantages and disadvantages, and so is worth considering as the community tackles how to maximally extract small scale information from the next generation of redshift surveys.

\subsection{Required developments}

Before a method like the one we have described here can be applied to real observational data, there is some additional work that is required. We outline some of these developments now.

\subsubsection{Variations in the galaxy-halo connection}

On sufficiently large scales, galaxies are expected to be almost unbiased tracers of the cosmic velocity field \citep{2018ApJ...861...58C}. However, effects like dynamical friction (which affects dark matter halos, but not individual dark matter particles) can lead to halos (and hence galaxies) having velocity distributions that differ from that of the matter distribution on smaller scales \citep[e.g.][]{2018MNRAS.474.3746A}.

In this work we used (sub)halos as our galaxies, which should produce a more realistic velocity bias than had we simply used dark matter particles. However, exactly which halos get populated with galaxies could potentially alter the velocity bias, leading to different GIK parameters for different galaxy samples. This means that in order to apply our method to real data, it would be important to populate the simulations -- used to train the GIK parameter emulator -- with galaxies in a manner that is close to how the real Universe populated its dark matter halos with galaxies.

Exactly which galaxies are selected in some survey (and therefore how those galaxies populate the underlying dark matter structures) is expected to lead to differences in their kinematics. For example, \citet{2018MNRAS.475.2530O} found that populating a DM-only simulation with galaxies using a semi-analytic model lead to different kinematics for stellar mass selected galaxies and galaxies with prominent emission lines. As a simple example, if part of what quenches star formation in galaxies and leads them to turn from blue to red is environmental effects, one expects some blue galaxies that fall into a cluster will be made red once inside. Within the GIK model this would produce quite different $\fvir(r)$ for red and blue galaxies, with (at fixed radius) blue galaxies more likely to be infalling, and red galaxies more likely to be virialised within the cluster potential.

Potentially better than trying to put galaxies into the simulations in the ``correct'' way, would be to use a parameterised model to populate the simulations with galaxies, populating each simulation numerous times using different galaxy--halo connection parameters. The GIK parameter emulator could then be built to cover variations in both cosmological and galaxy--halo connection parameters. Then, the galaxy--halo connection parameters could be left free in the fit, and marginalised over when drawing conclusions about cosmological parameters. This sort of approach is what is done when directly emulating redshift-space clustering data vectors from simulations \citep{2019ApJ...874...95Z, 2022MNRAS.515..871Y, 2023MNRAS.520.5373L, 2023arXiv230212379K}. In our case it would be non-trivial, because fitting for the GIK parameters is a reasonably computationally expensive step, so doing this many times over with different galaxy--halo connection parameters for each simulated cosmology would require a considerable amount of computing time. Nevertheless, such an endeavour would be important to properly incorporate how uncertainty on the way galaxies trace dark matter leads to uncertainties on the GIK parameters for a given cosmology.


\subsubsection{Prescription for the real space clustering}
\label{sect:real_space_clustering}

Throughout this paper we have assumed that the real space clustering is known, and that we only need to model the effects of RSD on this to get $\xicg$. When applying this method to real data, the real space clustering will not be known a priori, and must either be measured or modelled. Measuring the real space clustering could be done by measuring the projected correlation function (usually denoted $w_\mrm{cg}(\theta)$, which expresses the cross-correlation between the positions-on-the-sky of clusters and galaxies) and applying an inverse Abel transform \citep[e.g.][]{2019RScI...90f5115H} to get $\xi_\mrm{cg}^r(r)$. Modelling $\xi_\mrm{cg}^r(r)$ could involve using an emulator for the real-space clustering's dependence on cosmology \citep[e.g.][]{2023MNRAS.523.3219C}, or just using a flexible parametric model for $\xi_\mrm{cg}^r(r)$ and adding $w_\mrm{cg}(\theta)$ to the data vector being fit to, in order to help constrain this model's parameters.

We note that inferring the real space clustering has some associated challenges. For example, the projected clustering 
\begin{equation}
w_\mathrm{cg}(r_p) = \int_{- r_{\pi,\mathrm{max}}}^{r_{\pi,\mathrm{max}}} \xicg(r_p, r_\pi)
\end{equation}
integrates over the line-of-sight, removing any dependence on RSD. However, this is only exactly true for infinite $r_{\pi,\mathrm{max}}$, and \citet{2022MNRAS.517..374H} (when fitting models to the galaxy-group cross-correlation from the GAMA survey) found that a large enough $r_{\pi,\mathrm{max}}$ to achieve results that were independent of RSD effects, lead to measurements that were too noisy to be useful.

In addition, the projected clustering faces observational challenges associated with how galaxy spectra are obtained. For example, both SDSS and now DESI cannot obtain spectra for sufficiently nearby pairs of galaxies without multiple observations of the same field \citep{2017MNRAS.467.1940H, 2019MNRAS.484.1285S}. The bias to the measured correlation function that would come from preferentially missing galaxy pairs with small separations can be corrected for \citep[e.g.][]{2017MNRAS.472.1106B}, but these observational systematic effects nevertheless increase the uncertainties on small-scale clustering. This means that our inferred real space clustering may have significant uncertainty, which would translate into increased uncertainty on the velocity distribution (and hence the cosmological parameters) when fitting the GIK model to $\xicg$.

\subsubsection{Distribution of cluster masses}

At present we measure the GIK parameters from simulations for multiple narrow cluster mass bins, and use this to build an emulator for $p(v_r, v_t | r, M)$ (the cluster-galaxy pairwise velocity distribution, for a given cluster-galaxy separation and cluster mass). A realistic cluster sample, selected on a property like richness or X-ray luminosity, would be expected to have a sizeable range of cluster masses. If we have a sample of clusters with a number density per unit cluster mass of $\mrm{d}n/\mrm{d} M$, then one might imagine that we can find the GIK parameters appropriate for this sample of clusters as some sort of weighted average of the GIK parameters for different cluster masses. If we take equation~\eqref{eq:xicg_double_integral} and make explicit the dependence of various quantities on the mass of the galaxy clusters ($M$), we have at a particular mass that
\begin{multline}
1 + \xicg(\rp, \rpi | M) = \frac{H(z)}{1+z} \times \\ \int \frac{1 + \xi_\mathrm{cg}^r(r | M)}{\cos \theta} \int p\left( v_r, v_t | r, M \right) \, \mathrm{d}v_r \, \mathrm{d}y.
\label{eq:xicg_double_integral_M200}
\end{multline}
The average redshift-space cross-correlation with galaxies for our sample of clusters is 
\begin{equation}
\left< \xicg(\rp, \rpi) \right> = \int \xicg(\rp, \rpi | M) p(M) \, \mrm{d}M,
\label{eq:mean_redshift_space_clustering}
\end{equation}
where $p(M) = \frac{\mrm{d}n}{\mrm{d} M} / \int \frac{\mrm{d}n}{\mrm{d} M} \mrm{d} M$ is the normalised probability density associated with the cluster mass distribution. Similarly, the real-space clustering of our cluster sample is 
\begin{equation}
\left< \xi_\mathrm{cg}^r(r) \right> = \int \xi_\mathrm{cg}^r(r | M) p(M) \, \mrm{d}M.
\end{equation}
Putting equation~\ref{eq:xicg_double_integral_M200} into equation~\ref{eq:mean_redshift_space_clustering} we get
\begin{multline}
1 + \left< \xicg(\rp, \rpi) \right> = \frac{H(z)}{1+z} \times \int p(M) \\ \left[ \int \frac{1 + \xi_\mathrm{cg}^r(r | M)}{\cos \theta} \int p\left( v_r, v_t | r, M \right) \, \mathrm{d}v_r \, \mathrm{d}y \right] \, \mrm{d}M.
\end{multline}
The form of this integral means that 
we cannot simply calculate $\left< \xicg(\rp, \rpi) \right>$ using $\left< \xi_\mathrm{cg}^r(r) \right>$ and some appropriately-averaged $\left< p( v_r, v_t | r) \right>$. To properly calculate a model $\left< \xicg(\rp, \rpi) \right>$ for a broad cluster mass distribution we would need to have a model for $\xi_\mathrm{cg}^r(r | M)$, a model for $p\left( v_r, v_t | r, M \right)$ (which our emulator for the GIK parameters provides), and a model for $p(M)$.

This is problematic for the approach (described in Section~\ref{sect:real_space_clustering}) of using the observed projected clustering to infer $\xi_\mathrm{cg}^r(r)$, as this would produce a measurement of $\left< \xi_\mathrm{cg}^r(r) \right>$ -- averaged over the cluster sample -- without knowing about its mass dependence. For cluster mass distributions that are reasonably narrow, it may be sufficiently accurate to combine some sort of mass-averaged  $p\left( v_r, v_t | r \right)$ with $\left< \xi_\mathrm{cg}^r(r) \right>$. Generating cluster samples with suitably narrow mass distributions will be helped by the ongoing development of low-scatter cluster mass proxies \citep[e.g.][]{2020OJAp....3E..12E}, but we leave an investigation of how well this would work, and how one would best average $p\left( v_r, v_t | r, M \right)$, to future work. 

If one instead takes the approach of having a parameterised model for $\xi_\mathrm{cg}^r(r)$ and including $w(\theta)$ in the data vector being fit to, then this could be extended to have a parameterised model for $\xi_\mathrm{cg}^r(r|M)$ as well as $p(M)$. Combining these with $p\left( v_r, v_t | r, M \right)$ from the GIK parameter emulator we can then calculate both the redshift-space and projected correlation functions, allowing the parameters of the $\xi_\mathrm{cg}^r(r|M)$ and $p(M)$ models to be constrained by the clustering data.

\subsubsection{Extension to higher redshifts}

In this work our GIK parameter emulator was built only using simulation data from $z=0$. Extending this to work at different redshifts would simply require using simulation outputs at different redshifts. We could either build an emulator for the specific effective redshift of some target survey, or we could build the emulator with redshift being an additional input parameter along with cosmology and cluster mass, and use simulation outputs at many different redshifts in the GIK parameter training data. This emulator could then be applied to data from any redshift (including clustering data in different redshift bins).

\subsubsection{Selection effects}

Our model for redshift-space clustering assumes that the real-space clustering is isotropic, and that RSD due to cluster/galaxy velocities breaks this. However, the selection of our clusters and/or galaxies may break this assumption of isotropy. For example, the processes involved in optical cluster selection typically lead to the over-densities selected as clusters being preferentially elongated along the line-of-sight \citep{2014MNRAS.443.1713D}. If the cluster shape is correlated with the distribution of galaxies around it, then this would lead to the real-space clustering of galaxies around clusters already being anisotropic (with respect to the line-of-sight to the cluster) before accounting for RSD. 

Assessing the potential impact of this, as well as testing any mitigation strategies, would most-likely require realistic mock data to which cluster selection algorithms could be applied and a mock analyses performed. For example, the Cardinal mock galaxy catalogues \citep{2023arXiv230312104T} accurately reproduce the abundance of galaxy clusters, and could be used to assess how selection effects (especially orientation-dependent ones) might lead to biases in the inferred cosmological parameters.

\subsubsection{Artificial disruption in simulations}

Our approach relies on $N$-body simulations faithfully reproducing the cluster--galaxy pairwise velocity distribution. There has been recent literature about artificial disruption/destruction of subhalos in simulations \citep[e.g.][]{2018MNRAS.474.3043V, 2018MNRAS.475.4066V, 2021MNRAS.503.4075G, 2023arXiv230500993D}, with claims that a significant fraction of subhaloes that should survive are destroyed due to numerical effects. If the subhalos artificially destroyed preferentially have certain orbits, their destruction would bias the inferred GIK parameters, although this effect should primarily be on small scales (within approximately the cluster virial radii) where the GIK model cannot currently be applied to infer unbiased cosmological constraints anyway.

Fully understanding the potential impact of this would require a dedicated simulation study. Here, we note that the relatively good agreement in Fig.~\ref{fig:TNG_ZW_Fig5} between the GIK functions for TNG300, and a simulation with 64 times worse mass resolution, is a cause for optimism. That said, \citet{2018MNRAS.475.4066V} warn that because of the way in which both softening and particle mass are typically varied as simulation resolution is changed, it is possible for results to appear converged while not having converged to the true (infinite resolution) result.

\subsubsection{The effects of simulation box size}

As previously stated, a key advantage of our method is that it does not suffer too much from the noise associated with relatively small simulation volumes (see Section~\ref{sect:pros_cons_vs_direct_xicg}), enabling the use of small simulations to build an emulator for the pairwise velocities. However, small volume simulations do not just provide noisier estimates of quantities compared with their larger box-size counterparts, but can produce biased estimates. This arises if large-scale density fluctuations (not present in smaller volume simulations) impact the quantities of interest.

Large scale power is certainly important for the distribution of the peculiar velocities of dark matter halos \citep[see, for example, Fig.~9 of][]{1997MNRAS.286...38C}. However, it is not clear whether the pairwise velocities of halos/subhalos with separations much smaller than the box-size are appreciably impacted by the finite box-size and the associated absence of large-scale power. \citet{2023MNRAS.525.1039M} recently presented results for the pairwise velocities of the matter field, from ``scale free'' simulations, which have power-law power spectra. Box size effects were found to be strongly dependent on the spectral index of the power-spectrum in these simulations, making it hard to extrapolate to the case of $\Lambda$CDM-like universes, in which the power-spectrum is not scale free. We are not aware of work looking at this in a $\Lambda$CDM context, but a future measurement of the pairwise velocity distribution as a function of box-size is one of the steps that should be taken before a method like ours is applied to real data, in order to demonstrate the level at which the GIK parameters inferred from simulations might be biased by the box-size of the simulations used.

\subsection{Null tests of General Relativity}

Here, we have built an emulator for a specific theory of modified gravity, $f(R)$. This is just one of many alternatives to GR, and tying our modelling procedure to a particular modified gravity theory is somewhat undesirable. To understand alternative strategies to test theories of gravity, it is instructive to consider the case of RSD on larger scales. On large scales we can make use of linear theory, for which the matter power spectrum in redshift space, $P_\mathrm{m}^s(k, \mu)$, is given by \citep{1987MNRAS.227....1K}
\begin{equation}
P_\mathrm{m}^s(k, \mu) = (1 + f \mu^2)^2 P_\mrm{m}^r(k),
\end{equation}
where $P_\mrm{m}^r(k)$  is the (isotropic) real-space power spectrum, $\mu \equiv \hat{\vect{k}} \cdot \hat{\vect{z}}$ is the cosine of the angle between wavevector $\vect{k}$ and the line-of-sight direction, and $f$ is the linear growth rate \citep[for a definition, see e.g.][]{2001MNRAS.322..419H}. Within a $\Lambda$CDM + GR universe, the value of $f$ is closely approximated by $f \approx \Omega_\mrm{m}^{0.55}$ \citep{2005PhRvD..72d3529L}. Nevertheless, when fitting to observational data, $f$ and $\Omega_\mrm{m}$ can be treated as independent parameters. The inferred values of $f$ and $\Omega_\mrm{m}$ can then be compared, to see if the relationship between them is that expected in a $\Lambda$CDM + GR universe, or if they appear inconsistent with this. Treating the standard cosmological model as some sort of \emph{null hypothesis}, and then seeing if there is evidence for deviations to this, is attractive because it does not tie the analysis to any particular alternative, but just seeks to investigate whether there is evidence for some departure from $\Lambda$CDM + GR.

On the face of it, our method for modelling $\xicg$ is not so amenable to this style of analysis, because the emulator is essentially interpolating between a set of self-consistent simulations, and so it cannot produce predictions that are not self-consistent (such as a $\Lambda$CDM + GR universe with $f \not\approx \Omega_\mrm{m}^{0.55}$). It is interesting to consider how we could extend our modelling procedure to be more analogous to the case described above (where $f$ and $\Omega_\mrm{m}$ are treated as being independent when modelling RSD on linear scales). One possibility would be to treat our RSD model primarily as a way to measure the mass of the cluster sample, assuming GR within the model. We could build an emulator using $\Lambda$CDM + GR simulations (which, no longer having a variable $\fr0$, would no longer suffer the $M_{200}$--$\fr0$ degeneracy that we see in our results). We could then use this emulator to fit to an observed data vector. Comparing the mass inferred from fitting to $\xicg$ with that inferred using an independent method (with weak lensing being an obvious choice), we could then see if these two masses agree or not, with a disagreement hinting towards non-GR gravity.

Another alternative, would be to again build an emulator for the GIK parameters using $\Lambda$CDM + GR simulations, but then introduce one or more re-scaling parameters that would shift the GIK function curves. For example, one could imagine a single re-scaling parameter, $f_\mrm{vel}$, such that $v_\mrm{r,c} \to f_\mrm{vel} \, v_\mrm{r,c}$, $\sigma_\mrm{rad} \to f_\mrm{vel} \, \sigma_\mrm{rad}$ and $\sigma_\mrm{tan} \to f_\mrm{vel} \, \sigma_\mrm{tan}$. Fitting for the cosmological parameters alongside $f_\mrm{vel}$, evidence that $f_\mrm{vel} \neq 1$ would point towards some sort of modification to gravity.

\section{Conclusions}
\label{sect:conclusions}

We have presented a method to model the cluster-galaxy cross-correlation function in redshift-space, $\xicg$, based upon a slightly modified version of the Galaxy Infall Kinematics (GIK) model from \zw. We demonstrated how to find the posterior distribution for the parameters of the GIK model from 3D simulation data, with the data required being both positions and velocities of clusters and galaxies extracted from $N$-body simulations. Repeating this procedure for simulations run with different cosmologies (from the FORGE simulations, that simulate different $f(R)$ modified gravity cosmologies), we built up training data for the cosmology-dependence of the GIK parameters.

We used this training data to build Gaussian Process emulators for how the different GIK parameters depend upon the cosmological parameters and the theory of gravity. Combining these GIK parameter emulators with knowledge of the real space cluster-galaxy cross-correlation, allowed us to calculate $\xicg$ as a function of cosmological parameters. Comparing these model predictions with an observed data vector, we can then make inferences about the cosmological parameters.

The finite volume of the simulations, leads to uncertainties on our training data. We presented a method to propagate uncertainty on the GIK parameters from the training data, through to uncertainty on the model data vector. This included a procedure that we call ``GIK parameter diagonalisation'', a linear transform of the GIK parameter vector that allows separate Gaussian Processes (one for each ``diagonalised'' GIK parameter) to properly account for the covariance between uncertainties on the different GIK parameters in the training data.

Fitting our model to mock data generated by the model itself, we showed forecasts for the cosmological constraining power that RSD in the cluster-galaxy correlation function on small scales will provide with data from a survey like DESI. We found that a combination of the typical mass of our cluster sample and the strength of modifications to gravity was tightly constrained, such that an independent measurement of cluster mass (which could be provided by weak gravitational lensing) will enable RSD to tightly constrain theories of gravity.

Finally, we applied our method to mock data extracted directly from $N$-body simulations. We found that our procedure leads to a biased inference on the cosmological parameters if fitting the correlation function all the way down to a scale of $1 \, \hmpc$. However, these biases are strongly reduced when cutting out the smallest scales ($< 3 \, \hmpc$), at the expense of broadened constraints on the cosmological parameters. We attribute these biases to deficiencies in the model on a scale of $1-3 \, \hmpc$, with these deficiencies visible in both Fig.~\ref{fig:Pvrvt_example} and Fig.~\ref{fig:GIK_vs_sim}.

A large fraction of the cosmological constraining power from current and future redshift surveys comes from small scales, which will most likely require modelling methods based around the use of $N$-body simulations. Our approach -- of using simulations to build an emulator for the RSD kernel, which can be combined with some other model for the real-space clustering -- is quite different from other approaches in the literature that directly measure the redshift-space correlation function from simulations \citep{2019ApJ...874...95Z, 2022MNRAS.515..871Y, 2023MNRAS.520.5373L, 2023arXiv230212379K}. Advantages of our approach include:
\begin{itemize}
\item It separates out the two conceptually different components that produce the redshift-space clustering: the real-space clustering and the RSD kernel. This allows developments to be made to either component independently, and can also help with understanding and intuition.
\item It allows simulations with substantially smaller volumes than observational surveys to be used in the modelling of those large surveys, because imposing a functional form for the RSD kernel reduces the impact of noise in the simulations (see Fig.~\ref{fig:GIK_vs_sim}). This means that more simulations can be run, covering a wider range of possible cosmologies.
\item It uses simulations to predict the ingredient of small-scale redshift-space clustering that they appear to predict more robustly. As discussed in Appendix~\ref{App:TNG_convergence}, for the TNG300 simulations: the RSD kernel's contribution to $\xicg$ is better converged with respect to numerical resolution than the real space clustering's contribution to $\xicg$ is. 
\end{itemize}

The main drawback of our approach is that it relies on the GIK model being a good description of the true pairwise velocities. While this generally seems to be true, there is evidence that at radii $\lesssim 3 \, \hmpc$ the ``infalling'' component of the GIK model does not adequately describe the velocities from simulations (see the ``banana shaped'' $P(v_r, v_t)$ data and best-fit model in Fig.~\ref{fig:Pvrvt_example}). An avenue for future research would be to investigate modifications to the functional form for $P(v_r, v_t)$ that allow it to better describe the simulation data, with the goal being a model that could be applied down to very small scales without leading to biased inferences of the cosmological parameters.







\section*{Data Availability}

The Indra and IllustrisTNG simulation data is publicly available, while the FORGE simulation data can be obtained from the authors of \citet{2022MNRAS.515.4161A}. The derived data products underlying this article (including GIK parameter posteriors for the FORGE simulations, data covariance matrices calculated from Indra, and Gaussian Process kernel hyperparameters) will be shared on reasonable request to the corresponding author.

\section*{Acknowledgments}

The authors thank Christian Arnold for providing us with access to the FORGE simulations, and thank David Weinberg, Chun-Hao To, Peter Taylor, Alkistis Pourtsidou, Benjamin Bose and John Peacock for helpful discussions. This research was carried out at the Jet Propulsion Laboratory, California Institute of Technology, under a contract with the National Aeronautics and Space Administration (80NM0018D0004). The authors are grateful for funding from JPL's R\&TD 7x ``Future of Dark Sector Cosmology: Systematic Effects and Joint Analysis'' (01STRS/R.23.312.005). The High Performance Computing resources used in this investigation were provided by funding from the JPL Information and Technology Solutions Directorate.

This work made use of the following software packages: \href{https://github.com/astropy/astropy}{{Astropy}}
\citep{astropy1, astropy2},
\href{https://emcee.readthedocs.io/en/stable/}{{emcee}}
\citep{ForemanMackey:2013io},
\href{https://getdist.readthedocs.io/en/latest/intro.html}{{GetDist}}
\citep{2019arXiv191013970L},
\href{https://github.com/matplotlib/matplotlib}{{Matplotlib}}
\citep{matplotlib},
\href{https://github.com/numpy/numpy}{{NumPy}}
\citep{numpy},
and
\href{https://github.com/scipy/scipy}{{Scipy}}
\citep{scipy}.
This work used some of the \href{https://www.tng-project.org/data/docs/background/#sec4}{IllustrisTNG} simulations \citep{2018MNRAS.475..648P, 2018MNRAS.475..676S, 2018MNRAS.475..624N, 2018MNRAS.477.1206N, 2018MNRAS.480.5113M}. It also used the Indra simulations \citep{2021MNRAS.506.2659F}, accessing them via the SciServer science platform \citep{2020A&C....3300412T}.

\copyright 2023.    All rights reserved.

\bibliographystyle{mnras}

\bibliography{bibliography}

\appendix

\section{Functional form of the GIK functions}
\label{App:GIK_functions}

As discussed in Section~\ref{sect:GIKparams}, the variation of the GIK functions with radius is described by the set of GIK parameters. In Table~\ref{Tab:GIKfunc} we provide the parametric forms for the GIK functions.
We note that:
\begin{itemize}
\item The functional forms for $\sr$, $\st$, $\dof$ and $\alpha$ are all the same as one another (and the same as that used by \zw for these GIK functions), but we parameterise them slightly differently based on our experience of parameter degeneracies when fitting these functions to simulation data.
\item For $\fvir$ we use the same functional form as in \zw, but we fix $\gamma=2$ rather than $\gamma = 3$.
\item We use a different functional form for $\sv$ from that in \zw. This was because the form listed in \zw eq.~(9) did not seem to agree with what was plotted in \zw Fig.~6. The functional form we used was inspired by \citet{2022arXiv220413131A}.
\end{itemize}

\begin{table*}
\input{GIK_param_table}

\caption{The functional forms for the various GIK functions, with the corresponding free parameters (the ``GIK paramaters'') listed. A GIK parameter vector, $\vect{g}$, is the set of 24 dimensionless numbers listed in the ``GIK parameters'' column, working from the top down. Our prior on $\vect{g}$ (used when measuring the GIK parameters directly from position/velocity data from simulations) is a separable function of each GIK parameter, and is a uniform prior in each case, with the upper and lower limits enclosed in square brackets in the ``Prior'' column.}
\label{Tab:GIKfunc}
\end{table*}

\section{Alternative likelihood for fitting GIK parameters to a simulation}
\label{App:no_binning_likelihood}

\begin{figure}
        \centering
        \includegraphics[width=\columnwidth]{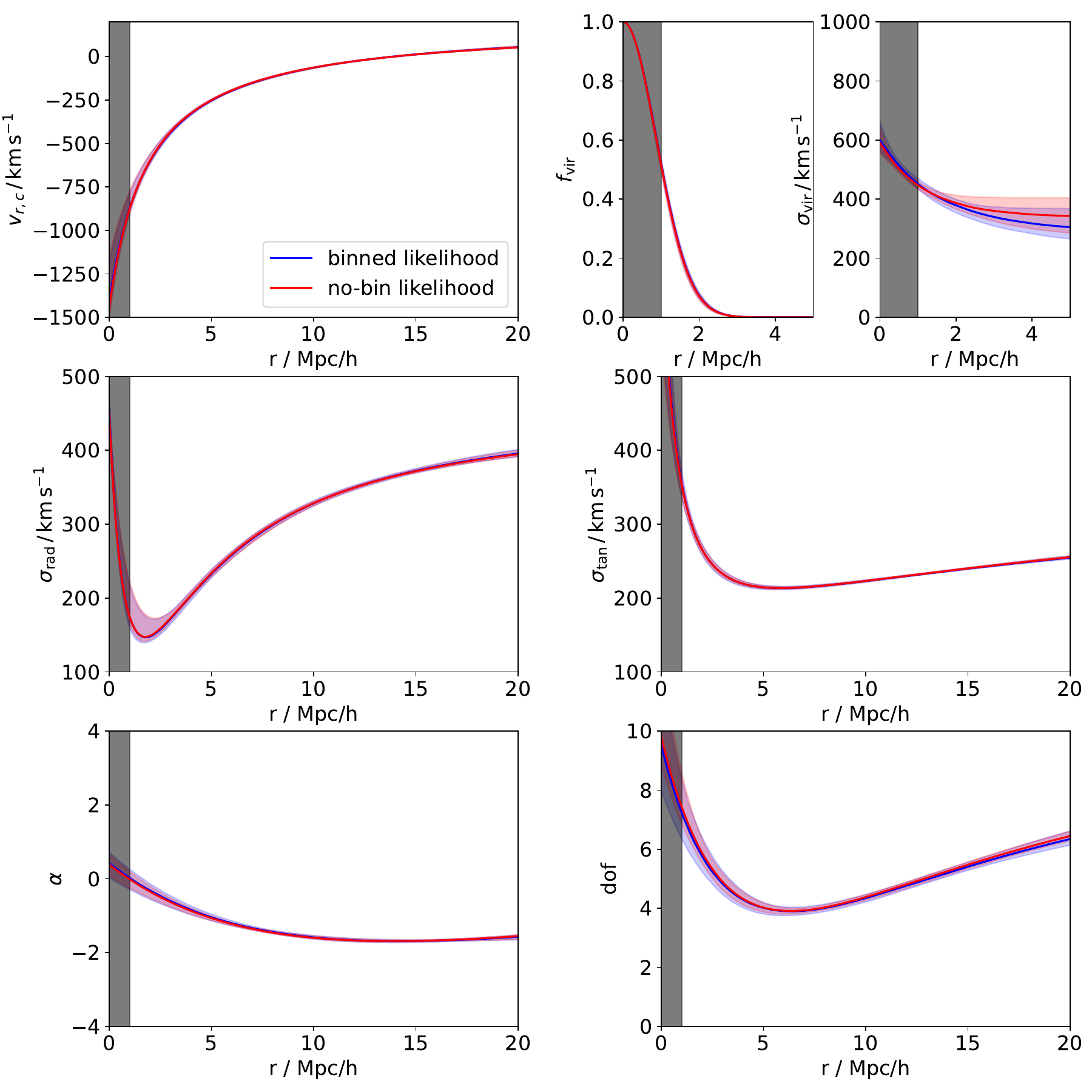}     
	\caption{The GIK function posteriors, from fitting for the GIK parameters using the binned and no-binned likelihoods described in Appendix~\ref{App:no_binning_likelihood}. The solid lines show the median GIK functions as a function of radius, while the shaded regions show the $2 \sigma$ uncertainty (the 2.5th to 97.5th percentiles of the GIK functions at each radius).}
	\label{fig:no-bin_likelihood}
\end{figure}

For our fiducial Indra simulation volume (for which $\xicg$ is plotted in Fig.~\ref{fig:example_xi_cg}), we tested whether our inferred GIK parameter vector, $\vect{g}$, was different if we used the likelihoods defined in eq.~\eqref{eq:Pvrvt_likelihood} or eq.~\eqref{eq:likelihood_no_binning}. For the first of these, the data that we fit to is calculated by binning cluster-galaxy pairs into radial shells and then within each radial shell further binning them in velocity ($v_r$, $v_t$) space. We then used MCMC to fit $\vect{g}$ to these binned counts, assuming that the counts obey Poisson statistics, with an expectation value for each pixel that depends on $\vect{g}$. For the second likelihood, the data is simply the set of $r$, $v_r$ and $v_t$ for each cluster-galaxy pair, with the likelihood being the product over all cluster-galaxy pairs of the probability density $P(v_r, v_t | r)$.

The results of these MCMCs are chains of samples drawn from the posteriors on the GIK parameters, which we can map into draws from the posteriors on the GIK functions by evaluating the functions at the locations of the GIK parameter samples. In Fig.~\ref{fig:no-bin_likelihood} we plot these GIK function posteriors, showing the median GIK function values as well as the $2 \sigma$ uncertainties. The results are virtually indistinguishable between using the two different likelihoods.

In terms of how important the small differences in Fig.~\ref{fig:no-bin_likelihood} are to our use case of using the GIK model to make model predictions for $\xicg$, we performed the following test. Keeping the real-space clustering fixed, we used the maximum-likelihood GIK parameters from both the binned and no-binned likelihood approaches to calculate $\xicg$. The difference between these two $\xicg$ model data vectors, when expressed as a $\chi^2$ using the data covariance matrix described in Section~\ref{sect:cov}, is 0.16, very small compared with the differences one gets when calculating $\xicg$ through the GIK model using Indra simulations run with the same cosmology but with different random initial conditions (see Section~\ref{sect:pros_cons_vs_direct_xicg}).

\section{The impact of simulation resolution and the inclusion of baryonic physics on the predicted $\xicg$}
\label{App:TNG_convergence}

In Section~\ref{sect:TNG_convergence} we presented the results for both the GIK functions and the real space clustering of different variants of the TNG300 simulation. Here we show the relative importance of these differences for the calculation of a model $\xicg$.

For each simulation, we measured the maximum likelihood GIK parameters, $\vect{g}$, as well as the real-space clustering, $\xi_\mathrm{cg}^r(r)$. We then used combinations of $\vect{g}$ and $\xi_\mathrm{cg}^r$ measured from different simulations to calculate various model $\xicg$. The similarity of two different $\xicg$, say $\xi_\mathrm{cg,1}^s$ and $\xi_\mathrm{cg,2}^s$, was then assessed in terms of $\chi^2 = \xi_\mathrm{cg,1}^s \, \vect{C}^{-1} \, \xi_\mathrm{cg,2}^s$, using the data covariance matrix, $\vect{C}$, described in Section~\ref{sect:cov} (re-scaled for the $\sim (300 \mpc)^3$ volume of a TNG300 simulation). The results of various combinations were as follows:
\begin{enumerate}[(a)] 
\item When keeping the real space clustering fixed to $\xi_\mathrm{cg}^r$ from TNG300, but using $\vect{g}$ from TNG300 or TNG300 low-res, the $\chi^2$ difference was 1.0.
\item When keeping $\vect{g}$ fixed to that from TNG300, but using $\xi_\mathrm{cg}^r$ from TNG300 or TNG300 low-res, the $\chi^2$ difference was 18.
\item When keeping the real space clustering fixed to $\xi_\mathrm{cg}^r$ from TNG300, but using $\vect{g}$ from TNG300 or $\vect{g}$ from TNG300 DMO, the $\chi^2$ difference was 0.12.
\item When keeping $\vect{g}$ fixed to that from TNG300, but using $\xi_\mathrm{cg}^r$ from TNG300 or TNG300 DMO, the $\chi^2$ difference was 1.7.
\end{enumerate}
Comparison of (a) and (b) with (c) and (d) shows that $\xicg$ is more affected by a factor 64 change in resolution than baryons vs no baryons. Comparing (a) with (b) we see that using low-resolution simulation GIK parameters leads to a better estimate of the high-resolution $\xicg$, than if using low-resolution simulation real space clustering. And comparing (c) with (d) tells us that the same is true when considering a hydrodynamical simulation versus a dark matter only one.

\section{What is the primary source of model misspecification error?}
\label{App:model_misspec}

The right panel of Fig.~\ref{fig:GIK_vs_sim} made clear that there is substantial model misspecification on small scales ($\lesssim 3 \, \hmpc$). This means that the 24 GIK parameters are not adequately describing the velocity distribution at small radii. There are two obvious culprits for this, one is that the 7 GIK functions do not adequately describe the pairwise velocity distribution at fixed radius, and the second is that the GIK parameters do not adequately describe the radial dependence of the GIK functions. Knowing the relevant importance of these sources of model misspecification should aid with future efforts to improve the GIK model.

To assess this, we calculated model $\xicg$ data vectors that did not make use of the GIK parameter description of how the GIK functions vary with radius. Instead, having fit for the GIK functions in many radial shells, we directly used the GIK functions to describe the velocity distribution at radii within the relevant shell. Specifically, the GIK functions used to calculate a model $\xicg$ were (over the radial range $[ r_i, r_{i+1}]$) the maximum likelihood set of GIK functions fit to the $\{v_r, v_t\}$ data for cluster-galaxy pairs with separation in the range $[ r_i, r_{i+1}]$. We used radial shells that were $1 \, \hmpc$ thick.

\begin{figure}
        \centering
        \includegraphics[width=\columnwidth]{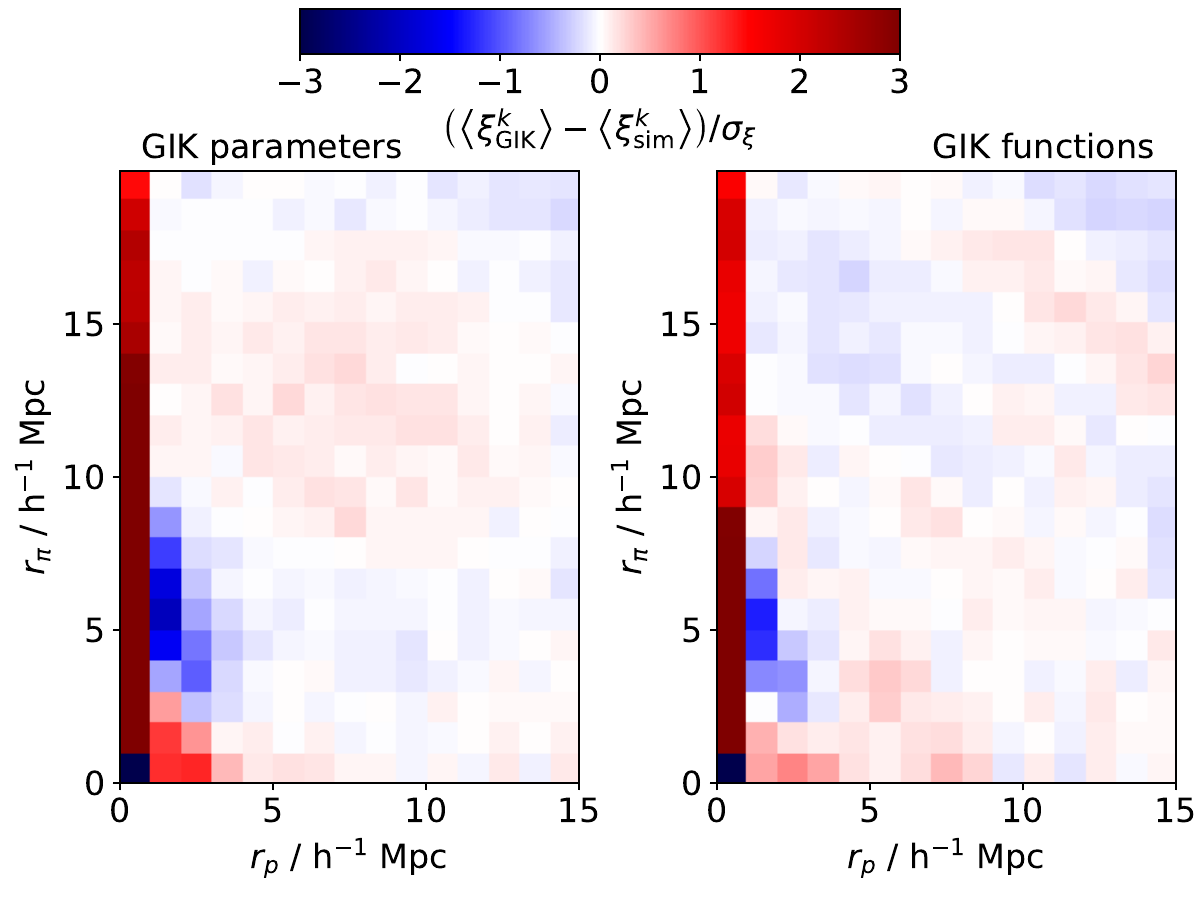}     
	\caption{Maps of the normalised residuals, similar to those in Fig.~\ref{fig:GIK_vs_sim}. Each panel shows the mean result over 24 Indra simulations. The mean $\xicg$ from the GIK models is compared with the mean $\xicg$ measured directly from the Indra simulations. In the left panel the model $\xicg$ is calculated in the usual way (where the velocity distribution at all radii is described by the 24 ``GIK parameters''). In the right panel the pairwise velocity distribution used for the model $\xicg$ treats each GIK function as constant within $1 \hmpc$ thick radial shells, with the values of the GIK functions within each radial shell obtained as the maximum likelihood values from fitting to the $\{v_r, v_t\}$ data in that shell.}
	\label{fig:GIKparam_vs_GIkfunc}
\end{figure}

We generated model $\xicg$ using both our fiducial approach (``GIK parameters'') and the approach just described (``GIK functions'') for each of 24 Indra simulations for which we had fit the GIK functions to the velocity data in spherical shells. Fig.~\ref{fig:GIKparam_vs_GIkfunc} shows the residuals between the mean of these 24 model $\xicg$, and the mean $\xicg$ measured directly from the Indra simulations, normalised by an error map, $\sigma_\xi$ (defined in Section~\ref{sect:reducing_noise}). The ``GIK functions'' residuals plot shows similar structure on small scales to the ``GIK parameters'' one, which means that a deficiency in the GIK functions is at least partly to blame for the mismatch between model and data $\xicg$ on small scales. Nevertheless, the residuals are less pronounced in the ``GIK functions'' case than the ``GIK parameters'' one, which means that the model could be improved somewhat without altering the 7 parameter model describing $P(v_r,v_t)$ at fixed radius, but simply by improving the description of how this varies with radius. Quantitatively, we found that the corresponding $\chi^2$ (using the covariance matrix from Section~\ref{sect:cov}, which ignores pixels in the lowest-$r_p$ bin) for the residuals plotted in Fig.~\ref{fig:GIKparam_vs_GIkfunc} were 23 and 11, for the ``GIK parameters'' and ``GIK functions'' cases respectively.

Achieving an improvement to the radial dependence of the GIK functions could be a case of devising better functional forms, that may then require more GIK parameters. An alternative, would be to use the Gaussian processes that currently model the cosmology dependence of the GIK parameters, to also model the radial dependence of the GIK functions. In essence, one would use a suite of simulations to measure the GIK functions at different points in cosmological parameter space, as well as for different cluster masses and different cluster-galaxy separations. Then a Gaussian process could be constructed for each of the GIK functions, that predicts the value of this GIK function as a function of cosmology, halo mass, and separation. Such a procedure could only improve the model so far though, given that the right panel of Fig.~\ref{fig:GIKparam_vs_GIkfunc} still shows residuals, that would require an improvement to the GIK functions themselves to remove. As noted in Section~\ref{sect:model_misspec}, the comparison between the data and model $P(v_r, v_t)$ in Fig.~\ref{fig:GIK_vs_sim} provides some guidance as to how the current GIK functions are deficient at small scales.

%

\bsp
\label{lastpage}

\end{document}

%% file: GIK_param_table.tex
\begin{tabular}{lllp{3cm}}
GIK function & GIK parameters & Prior & Notes on GIK function \\ \hline

$\sr(r) = q - qf \frac{r }{r_\mrm{min} } \left( \frac{\beta r_\mrm{min}}{r + (\beta - 1) r_\mrm{min} } \right)^\beta$ & \makecell[tl]{$\log_{10} q / \mathrm{km \, s^{-1}}$ \\ $f$ \\ $\log_{10} \beta$ \\  $\log_{10} r_\mathrm{min} / h^{-1} \mathrm{Mpc}$} & \makecell[tl]{$[2,3.5]$ \\ $[0,0.99]$ \\ $[0.04,1.7]$ \\ $[-0.5,2]$}  &  \\ \hline

$\st(r) = q - qf \frac{r }{r_\mrm{min} } \left( \frac{\beta r_\mrm{min}}{r + (\beta - 1) r_\mrm{min} } \right)^\beta$ & \makecell[tl]{$\log_{10} A / \mathrm{km \, s^{-1}}$ \\ $f$ \\ $\log_{10} \beta$ \\  $\log_{10} r_\mathrm{min} / h^{-1} \mathrm{Mpc}$} & \makecell[tl]{$[1.5,3]$ \\ $[0,0.99]$ \\ $[0.04,1.7]$ \\ $[-0.5,2]$} & $A$ is the minimum value of $\st(r)$, given by $A = q(1-f)$ \\ \hline

$\vrc(r) = q - t \left( \frac{r_i}{r+r_i} \right)^\beta$ & \makecell[tl]{$q / \mathrm{km \, s^{-1}}$ \\ $\log_{10} t / \mathrm{km \, s^{-1}}$ \\ $\beta$ \\  $\log_{10} r_i / h^{-1} \mathrm{Mpc}$} & \makecell[tl]{$[-1000,1000]$ \\ $[1,4]$ \\ $[0.1,4]$ \\ $[-1.5,2]$} &  \\ \hline

$\dof(r) = 1 + q - qf \frac{r }{r_\mrm{min} } \left( \frac{\beta r_\mrm{min}}{r + (\beta - 1) r_\mrm{min} } \right)^\beta$ & \makecell[tl]{$A$ \\ $f$ \\ $\log_{10} \beta$ \\  $\log_{10} r_\mathrm{min} / h^{-1} \mathrm{Mpc}$} & \makecell[tl]{$[0,10]$ \\ $[0,1]$ \\ $[0.1,4]$ \\ $[-0.5,2]$} & \makecell[tl]{$A = q(1-f)$ \\ $\dof \ge 1$} \\ \hline

$\alpha = q - qf \frac{r }{r_\mrm{min} } \left( \frac{\beta r_\mrm{min}}{r + (\beta - 1) r_\mrm{min} } \right)^\beta$ & \makecell[tl]{$q$ \\ $h$ \\ $\log_{10} \beta$ \\  $\log_{10} r_\mathrm{min} / h^{-1} \mathrm{Mpc}$} & \makecell[tl]{$[-3,10]$ \\ $[-10,10]$ \\ $[0.04,1.7]$ \\ $[-0.5,2]$} & $h = f \, q$ \\ \hline

$\fvir(r) = \exp \left( {-(r/r_0)^\gamma} \right)$ & $r_0 / h^{-1} \mathrm{Mpc}$ & $[0,10]$ & We fix $\gamma = 2$ \\ \hline

$\sv(r) = \sigma_0 \left( 1 + q \, \exp\left( {\frac{-r}{r_\mrm{vir}}} \right) \right)$ & \makecell[tl]{$\sigma_0 / \mathrm{km \, s^{-1}}$ \\ $q$ \\ $r_\mathrm{vir} / h^{-1} \mathrm{Mpc}$} & \makecell[tl]{$[100,2000]$ \\ $[0,2]$ \\ $[0,5]$} & Inspired by \citet{2022arXiv220413131A} 

\end{tabular}